\documentclass[pra,twocolumn,superscriptaddress,amsmath,amssymb]{revtex4-2}
\usepackage{amsmath}
\usepackage{amsfonts}
\usepackage{amssymb}
\usepackage{graphicx}
\usepackage{color}
\usepackage{txfonts}
\usepackage{float}
\usepackage[titletoc]{appendix}
\usepackage[colorlinks={true}]{hyperref}
\hypersetup{citecolor={blue}, filecolor={blue}, linkcolor={blue}, urlcolor={blue}}
\begin{document}


\title{\textbf{All-optical Raman control of ultracold atomic hyperfine states using pulsed
jump protocol} 
}%

\author{Xin-Xia Jian} 
\affiliation{Hunan Key Laboratory of Nanophotonics and Devices, Hunan Key Laboratory of Super-Microstructure and Ultrafast Process, School of Physics, Central South University,
Changsha 410083, China}

\author{Zhi-Jian Zheng} 
\affiliation{Hunan Key Laboratory of Nanophotonics and Devices, Hunan Key Laboratory of Super-Microstructure and Ultrafast Process, School of Physics, Central South University,
Changsha 410083, China}

\author{Jun-Jie Jiang} 
\affiliation{State Key Laboratory of Magnetic Resonance and Atomic and Molecular Physics, Innovation Academy for Precision Measurement Science
and Technology, Chinese Academy of Sciences-Wuhan National Laboratory for Optoelectronics, Wuhan 430071, China}

\author{Lin Zhou} 
\affiliation{State Key Laboratory of Magnetic Resonance and Atomic and Molecular Physics, Innovation Academy for Precision Measurement Science
and Technology, Chinese Academy of Sciences-Wuhan National Laboratory for Optoelectronics, Wuhan 430071, China}

\author{Chuan-Cun Shu} 
\email{cc.shu@csu.edu.cn}
\affiliation{Hunan Key Laboratory of Nanophotonics and Devices, Hunan Key Laboratory of Super-Microstructure and Ultrafast Process, School of Physics, Central South University,
Changsha 410083, China}

\author{Jun He} 
\email{junhe@csu.edu.cn}
\affiliation{Hunan Key Laboratory of Nanophotonics and Devices, Hunan Key Laboratory of Super-Microstructure and Ultrafast Process, School of Physics, Central South University,
Changsha 410083, China}


\date{\today}

\begin{abstract}
We develop a pulse-driven jump protocol to achieve all-optical Raman control of ultracold atomic hyperfine states. By establishing general conditions for adiabatic evolution between quantum states in parameter space, we derive the essential pulse area and phase conditions necessary for quantum state transfer in a resonant single-$\Lambda$ three-level system. We extend this approach to a double-$\Lambda$ four-level system by incorporating a neighboring intermediate state, which leads to a single-photon detuned $\Lambda$ three-level system. Through numerical simulations of the ultracold $^{87}$Rb atomic system, we demonstrate that high-fidelity and robust control of quantum state transfer can be achieved in the single-$\Lambda$ three-level system using stimulated Raman adiabatic passage (STIRAP) and the pulsed jump protocol. Furthermore, we show that the destructive quantum interference effects between resonant and detuned Raman pathways in the double-$\Lambda$ four-level system can be mitigated by optimizing the pulse area and two-photon detuning parameters within the pulsed jump protocol. This work presents a promising approach for achieving all-optical Raman control of quantum state transfer in ultracold atomic hyperfine states.
\end{abstract}
\date{\today}
\maketitle
\section{Introduction}
Due to their unique properties in long coherence times and a high degree of control, the hyperfine quantum states of ultracold atoms provide a highly controllable, coherent, and scalable platform for quantum computing \cite{Nielsen_book_2010,WuJinLei_PRApplied_2021,ZhangWeiYong_PRL_2023, Evered_nature_2023, CaoJiaHao_PRL_2023}, simulation \cite{Schäfer_NRP_2020, Meng_PRL_2023, Chalopin_PRL_2025, Halimeh_NP_2025}, and metrology \cite{ZhouLin_PRL_2015,ZhouLin_PRA_2021, Anders_PRL_2021,YinMoJuan_PRL_2022, Panda_NP_2024}. To manipulate transitions between hyperfine levels, shaped microwave pulses are typically used because its frequency matches the energy differences of the levels by carefully designing and controlling the microwave fields with controlled frequency, amplitude, and shape  \cite{Bohi_NP_2009,Riedel_Nature_2010,Pershin_(PRA)_2020,Kondo_PRA_2024}. However, employing microwave pulse techniques to control hyperfine energy levels in ultracold atoms presents several challenges and disadvantages. These stem primarily from decoherence effects, stringent precision requirements, and susceptibility to fluctuations in external fields.\\ \indent
Through the use of laser light to induce transitions between hyperfine states, all-optical Raman control of hyperfine energy levels in ultracold atoms is becoming increasingly prominent without the necessity of microwave pulses \cite{Condon_PRL_(2019),Arunkumar_PRL_2019, Chuan-Cun_PRA_2019, Weckesser_Nature_2021, Herbst_PRA_2022}. All-optical Raman control models typically involve an intermediate state in an excited electronic configuration to transfer the population between two hyperfine states within the lower (ground) electronic state \cite{RRS_PRL_(2019), Pucher_PRL_2024, Unnikrishnan_PRL_2024}, which holds potential for implementing quantum memory schemes \cite{Lvovsky_NaturePhotonics_2009, Heshami_JOMP_2016,Wolfowicz_book_2016}. This results in a standard quantum control scheme in a three-level \(\Lambda\) system for precisely manipulating the internal states of ultra-cold atoms. Based on the Raman process and the adiabatic approximation, a well-known technique called stimulated Raman adiabatic passage (STIRAP) is used to achieve coherent transfer between hyperfine quantum states \cite{Gaubatz_JCP_1990, ShuChuan-Cun_PRA_2009, Bergmann_JCP_2015, Vitanov_RMP_2017, Maddox_PRL_2024}. This technique employs a counterintuitive timing of the pump and Stokes laser pulses to couple the initial energy level with an intermediate level and then the intermediate level with the target level. By carefully controlling these pulses' intensity and phase relationships, the system can be effectively transferred from the initial energy level to the target level with high fidelity while minimizing the population at the intermediate level. The success of STIRAP relies on the adiabatic condition \cite{Bergmann_JCP_2015, Vitanov_RMP_2017, Maddox_PRL_2024,Messiah1965QuantumMV, Peterson_PRA_(1985), Kuklinski_PRA_(1989)}, which requires that the system's evolution occurs slowly compared to the inverse of the energy gap between the dark and bright states.\\ \indent 
Recent studies on the necessary and sufficient conditions for quantum adiabatic evolution reveal that the maintenance of adiabatic evolution is more closely related to a smoothly varying eigenpath than to the slow variation of eigenenergies \cite{NASC_PRA_(2016),Marzlin_PRL_2004,Tong_PRL_2007,Boixo_PRA_2010}. This insight has led to the development of quantum jump protocol \cite{ZhengWen_PRAppl_(2022),WangZhen-Yu_arXiv_(2023),GongMusang_PRA_(2023),WangZhen-Yu_PRA_2024}, which utilizes discrete jumps along the evolution path of the control parameters to implement quantum adiabatic processes. By employing a series of coherent pulses or rapidly changing fields parameterized by the adiabatic path, achieving adiabatic evolution while accumulating geometric phases in a significantly shorter timeframe for a given average energy is possible. This approach enables rapid evolution and can even circumvent path points where the Hamiltonian's eigenstates are not experimentally accessible \cite{XuKebiao_ScienceAdvances_(2019),WangZhen-Yu_arXiv_(2023)}. However, implementing the quantum jump protocol in ultracold atomic systems presents several challenges. Firstly, the pulsed control fields that are utilized to control the quantum system's evolution are expected to be switched on and off smoothly \cite{Harutyunyan_PRL_(2023),Stanchev_PRA_2024}, rather than being abruptly turned on and off with rectangular pulses \cite{Gevorgyan_PRA_(2021),Dridi_PRA_(2020),ZhangCheng_PRA_(2024),XiaYan_PRA_(2024)}.  
Secondly, the energy gaps between hyperfine levels are tiny  typically in GHz regimes, which limits the bandwidths of the pump and Stokes pulses to avoid potential mutual interactions between pulses \cite{Mahana_PRA_2024}. Thirdly, neighboring energy levels surrounding the intermediate state inevitably affect the adiabatic evolution \cite{Sierant_PRA_2023}.\\ \indent
This work explores all-optical quantum control of ultracold atomic hyperfine states via pulse-driven Raman transitions. We derive general conditions for adiabatic evolution by defining the trajectory of evolution within the parameter space. We introduce the fundamental mechanism of STIRAP and establish the pulse area and phase conditions for the pulse-driven jump protocol in a resonant single-$\Lambda$ three-level system. Furthermore, we extend our analysis to a double-$\Lambda$ four-level system by considering a neighboring energy level above the intermediate state.
To demonstrate our method, we examine a resonant single-$\Lambda$ three-level model in ultracold $^{87}$Rb atoms to assess the bandwidth applicability of the STIRAP and jump protocol scheme. We achieve complete population inversion with minimal transient population in the intermediate state. The protocol's robustness to pulse area errors improves as the number of pulse pairs increases. Additionally, we investigate the impact of a neighboring intermediate state on the pulse-driven jump protocol. We find that high-fidelity quantum state transfer can also be achieved in the four-level system by modulating the pulse area and two-photon detuning parameters. This work offers a thorough understanding of adiabatic evolution for quantum control of double-$\Lambda$ four-level systems, providing valuable insights for optimizing pulse-driven jump protocols in ultracold atomic systems.\\ \indent 
The remainder of the paper is organized as follows: Section~\ref{II} presents the general condition for adiabatic evolution and describes the theoretical methods of stimulated Raman adiabatic passage and the pulsed jump protocol. In Sec.~\ref{III}, we conduct numerical simulations and discussions to apply this protocol to both the three-level system and the four-level double \(\Lambda\) system of ultracold \({}^{87}\mathrm{Rb}\) atoms. We conclude our results in Sec.~\ref{IV}.
\section{Theoretical model and methods }
\label{II}
\subsection{General condition for adiabatic evolution}
\label{A}
To derive the general adiabatic condition, we consider a quantum system driven by a time-dependent Hamiltonian \(\hat{H}(t) \equiv \hat{H}(\vec{R}) \equiv \hat{H}(\theta)\), where \(\vec{R} = (R_1(\theta), R_2(\theta), \dots)\) is parametrized by the time-dependent monotonically increasing parameter \(\theta\). The evolution of the system from $t = t_0$ (with parameters $\vec{R} = \vec{R}_0$ and $\theta = \theta_0$) to $t = t_T$ (with $\vec{R} = \vec{R}_T$ and $\theta = \theta_T$) can be pictured as transporting a path $\vec{R}(\theta)$ in parameter space, with $\lambda(t) = \dot{\theta}(t)$ characterizing the speed of traversal along the path.
We denote the instantaneous orthonormal eigenstates and eigenenergies of $\hat{H}(\theta)$ by $\left\{\left| \psi_f(\theta) \right\rangle \right\}$ and $\left\{E_f(\theta)\right\}$, respectively.
The system dynamics driven by the Hamiltonian \(\hat{H}[\theta(t)]\) is fully determined by the corresponding unitary evolution propagator \(\hat{U}(t) = \hat{U}[\theta(t)]\), which can be decomposed as the product of gauge-invariant unitary operators
\begin{equation}
  \label{Eq:3}
\hat{U}\left( \theta \right) =\hat{U}_{\text{Adia}}(\theta )\hat{U}_{\text{Dia}}(\theta ),\end{equation}
where $\hat{U}_{\text{Adia}}(\theta )$ is an ideal quantum adiabatic evolution propagator and $\hat{U}_{\text{Dia}}(\theta )$ is a diabatic propagator that includes all the adiabatic errors. To achieve adiabatic evolution of the system, we need to eliminate all diabatic terms, that is, \(\hat{U}_{\mathrm{Dia}}(\theta) \) becomes an identity matrix. Consequently, the system's evolution propagator will follow the desired adiabatic evolution, i.e., \(\hat{U}(\theta)=\hat{U}_{\mathrm{Adia}}(\theta)\) .\\ \indent 
According to the result of Ref.  \cite{NASC_PRA_(2016)}, the nonadiabatic correction term $\hat{U}_{\text{Dia}}(\theta)$ satisfies the first-order differential equation $\frac{d}{d\theta} \hat{U}_{\text{Dia}}(\theta) = iW(\theta) \hat{U}_{\text{Dia}}(\theta)$ with the boundary condition $\hat{U}_{\text{Dia}}(\theta_0) = \mathbb{I} $, and the formal solution is $\hat{U}_{\text{Dia}}(\theta) = P \exp[{i\int_{\theta_{0}}^{\theta_{}} W(\theta^{\prime})]d\theta^{\prime}}$, where \( P \) denotes path ordering with respect to \( \theta \) in a manner analogous to time ordering. The generator \( W(\theta) \) is responsible for the nonadiabatic transitions,
\begin{equation}
  \label{Eq:5}
  W(\theta) = \sum_{f \neq h} Y_{f,h}(\theta) G_{f,h}(\theta),
\end{equation}
which can be decomposed into the dynamical phase accumulation functions 
\begin{equation}
 Y_{f,h}(\theta) = e^{i \left[ \alpha_h(\theta) - \alpha_f(\theta) \right]},
\end{equation}
determined by the eigenenergies with the dynamical phases
$ \alpha_f(\theta) = -\int_{\theta_0}^{\theta} {E_f(\theta^{\prime})}/{\lambda} d\theta^{\prime}$, and the geometric parts
\begin{equation}
 G_{f,h}(\theta) = e^{i \left[ \gamma_h(\theta) - \gamma_f(\theta) \right]} g_{f,h}(\theta)| \psi_f(\theta_0) \rangle \left\langle \psi_h(\theta_0) \right|,
\end{equation}
determined by the eigenstates consisting of the geometric functions $g_{f,h}(\theta) = \langle \psi_f(\theta) | i \frac{d}{d\theta} \left| \psi_h(\theta) \right\rangle$ and incorporating the phases \(\gamma_f(\theta) = \int_{\theta_0}^{\theta} \langle \psi_f(\theta') | i \frac{d}{d\theta'} | \psi_f(\theta') \rangle \, d\theta'\).
\\ \indent 
Based on this consideration,  the deviation from adiabatic evolution is described by $\hat{D}_{\text{Dia}}\equiv \hat{U}_{\text{Dia}}\left( \theta \right) -\mathbb{I},$ and the corresponding general upper bound of the uniform invariant norm for diabatic correction reads \cite{NASC_PRA_(2016)}
\begin{equation}
  \label{Eq:6}
\|\hat{D}_{\mathrm{Dia}}(\theta)\|\lesssim2(\theta-\theta_0)\sqrt{\epsilon G_{\mathrm{tot}}(G_{\mathrm{tot}}^2+G_{\mathrm{tot}}^{\prime})},
\end{equation}
where $G_{\mathrm{tot}}=\sum\limits_{f\neq h}\underset{\theta'\in \left[\theta _{0},\theta\right] }{\sup }\left\Vert G_{f,h}\left( \theta'\right) \right\Vert $ and $G_{\mathrm{tot}}^{\prime }=\sum\limits_{f\neq h}
\underset{\theta'\in \left[ \theta _{0},\theta \right] }{\sup }
\left\Vert \frac{d}{d\theta'}G_{f,h}\left( \theta' \right) \right\Vert $. The deviation from adiabaticity in Eq.~(\ref{Eq:6}) can be minimized by reducing the dynamic accumulation functions
\begin{equation}
  \label{Eq:7}
\epsilon_{f,h}(\theta) = \left|\int_{\theta_{0}}^{\theta}Y_{f,h}(\theta^{\prime})d\theta^{\prime}\right| < \epsilon \  (f\neq h),\end{equation}
with a fixed scale factor determined by the magnitude of $G_{\mathrm{tot}}$ and its derivative $G_{\mathrm{tot}}^{\prime }$,  i.e., $\epsilon$ approaching a value arbitrarily close to zero. It leads to that the system evolution is adiabatic along the entire finite path with $\hat{U}_{\mathrm{Dia}}(t)\to\mathbb{I}$. Under the premise that $F_{f,h}$ represent fast oscillating functions and the slowly varying functions $G_{f,h}$ are averaged out, with dynamical phase accumulation $\epsilon$ serving as the area quantization function, the condition $\epsilon \rightarrow 0$ is sufficient.
\subsection{Stimulated Raman adiabatic passage in a single $\Lambda$ system}
\begin{figure*} [ht]
  \centering
  \includegraphics[width=0.95\textwidth]{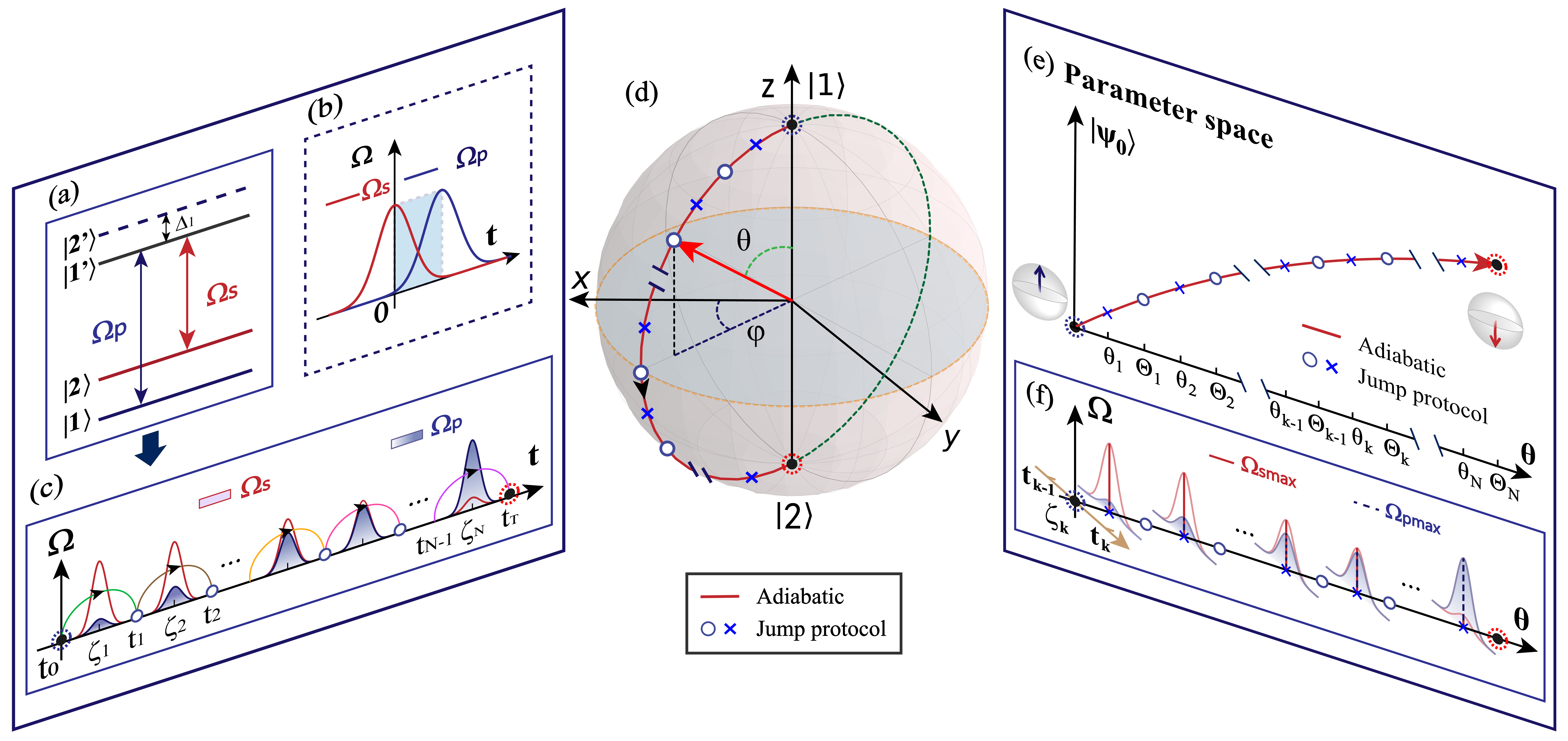}
  \caption{(a) Schematic of a four-level double-$\Lambda$ system. The hyperfine ground states $|1\rangle$, $|2\rangle$ are coupled to the excited states $|1'\rangle$ and $|2'\rangle$  via pump (blue) and Stokes (red) laser fields.
(b) Counterintuitive pulse sequence for STIRAP. The Stokes pulse (red solid line) precedes the pump pulse (blue solid line).
(c) Pulsed jump protocol in a three-level Raman system. Two synchronized double-pulse sequences drive stepwise population transfer between $|1\rangle$ and $|2\rangle$.
(d) Evolution trajectories on the Bloch sphere. STIRAP (red solid line) follows a smooth adiabatic path, while the pulsed jump protocol (blue circles: adiabatic points; blue crosses: non-adiabatic deviations) exhibits discrete jumps between states $|1\rangle$ and $|2\rangle$.
(e) State evolution pathways. Ideal adiabatic transfer (red solid line) and the pulsed jump protocol (blue circles and blue crosses) are compared.
(f) Parametric pulse amplitude modulation. The $k$-th pump pulse maximum versus Stokes maximum as functions of the mixing angle $\theta$. Pulse shapes (top axis) correspond to time intervals $(t_{k-1}, t_k)$.}
  \label{fig:0}
\end{figure*}
\label{STIRAP}
We now examine the adiabatic evolution via the STIRAP scheme in a $\Lambda$-type three-level system consisting of two low-lying levels, $\left| 1 \right\rangle$ and $\left| 2 \right\rangle$, with transition frequency $\omega_{21}$, and an intermediate level $|1'\rangle$, as illustrated in Fig.~\ref{fig:0}(a).
A pair of pump and Stokes pulses \(\mathcal{E}_p(t)=\mathcal{E}_{p0} f_p(t)\cos[\omega_{p}(t-t_p)]\) and \(\mathcal{E}_s(t)=\mathcal{E}_{s0} f_s(t)\cos[\omega_{s}(t-t_s)]\) with center frequencies \(\omega_p\) and \(\omega_s\), the strengths $\mathcal{E}_{p0}$ and $\mathcal{E}_{s0}$, the profile functions $f_p(t)$ and $f_s(t)$, and the center times $t_p$ and $t_s$, are used to resonantly couple the states \(\left| 1 \right\rangle\) and \(\left| 2 \right\rangle\) to the state \(\left| 1' \right\rangle\) with transition frequencies \(\omega_{1'1}\) and \(\omega_{1'2}\) and transition dipole moments \(\mu_{11'}\) and \(\mu_{21'}\). Within  the rotating wave approximation (RWA), the corresponding  Hamiltonian reads
\begin{equation}
  \label{Eq:8}
\hat{H}_{\text{RWA}}(t)=-\frac{1}{2}\left[ \Omega _{p}(t)e^{-i\varphi _{p}}\left\vert1\right\rangle \left\langle 1'\right\vert +\Omega _{s}(t)e^{-i\varphi_{s}}\left\vert 2\right\rangle \left\langle 1'\right\vert +\text{H.c.}\right],
 \end{equation}
where $\Omega_{p}(t)=\mu_{11'}\mathcal{E}_{p0} f_p(t)$ and $\Omega_{s}(t)=\mu_{21'}\mathcal{E}_{s0} f_s(t)$ denote the time-dependent Rabi frequencies of the pump and Stokes pulses, and the phases are $\varphi_{p}=\omega _{p}t_{p}$ and $\varphi_{s}=\omega _{s}t_{s}$. By diagonalizing the Hamiltonian $\hat{H}_{\text{RWA}}$ in Eq.~(\ref{Eq:8})
, we can obtain a time-dependent dark state
\begin{equation}
  \label{Eq:10}
\left\vert \psi _{0}(\theta(t) )\right\rangle =\cos \frac{\theta(t)}{2}\left\vert 1\right\rangle +e^{-i\varphi+i\pi}\sin\frac{\theta(t)}{2}\left\vert 2\right\rangle,
\end{equation}
with the eigenenergy $E_{0}=0$ and the relative phase $\varphi =\varphi_{s}-\varphi_{p}$. In Eq.~(\ref{Eq:10}), the mixing angle $\theta(t)$ is defined by $\tan\theta(t) /2= \Omega_{p}(t)/\Omega_{s}(t)$ that  changes monotonically over time with the initial condition $\theta_{0} = 0$. By defining  the effective  Rabi frequency  $\Omega_{\text{0}}(t)=\sqrt{\Omega _{p}^{2} (t)+\Omega _{s}^{2}(t)}$, the two Rabi frequencies $\Omega_{p}$ and $\Omega_{s}$ can be parameterized as $\Omega_{p} = \Omega_{\text{0}} \sin (\theta/2)$ and $\Omega_{s} = \Omega_{\text{0}} \cos (\theta/2)$.
By utilizing a pair of counterintuitive pump and Stokes pulses, the STIRAP can be established to maintain the system in a dark state, as described by Eq.~(\ref{Eq:10}) throughout the process. The evolution path of the dark state on the Bloch sphere is depicted by the red line in Fig.~\ref{fig:0}(d), assuming that the relative phase $\varphi$ remains constant over time, thereby preserving specific dynamic properties. The STIRAP approach typically requires sufficiently long pulse durations to ensure that the mixing angle $\theta(t)$ changes smoothly and continuously from $\theta_0=0$ to $\theta_T=\pi$, allowing the system to be transferred from the initial state $\left|1\right\rangle$ to the target state $\left|2\right\rangle$ with minimal occupation of the intermediate state $\left|1'\right\rangle$ along an adiabatic evolution path $\vec{R}(\theta)$. To speed up adiabatic processes with relatively short pulses, the pulsed jump protocol can transfer quantum states through a series of discrete, stepwise operations along the adiabatic evolution in a jump manner. 
\subsection{Pulsed jump protocol in a single $\Lambda$ system}
\label{D} 
We now explain how to design a pulse train comprised of a series of pump-Stokes pulse pairs to implement the pulsed jump protocol in the three-level system described above, which integrates the concepts of both the jump protocol \cite{GongMusang_PRA_(2023),WangZhen-Yu_arXiv_(2023)} and digital methods \cite{PAP1(2007), PAP2(2008), Harutyunyan_PRL_(2023)}.
 We consider $N$ pulse pairs that are separated from each other and do not overlap. The $k$-th pulse pairs can be defined by ($k=1,2,3\ldots,N $)
\begin{eqnarray} \label{pulse}
\begin{split}
\mathcal{E}_{pk}(t) &= \mathcal{E}_{pk} f_{pk}(t)\cos[\omega_{p}(t - \zeta_k)+\varphi _{pk}^0],\\
\mathcal{E}_{sk}(t) &= \mathcal{E}_{sk} f_{sk}(t)\cos[\omega_{s}(t - \zeta_k)+\varphi _{sk}^0],
\end{split}
\end{eqnarray}
where each subpulse interacts within the time interval \(t \in (t_{k-1}, t_k)\) with a duration of \(\tau = t_k - t_{k-1}\), and \(t_0 = 0\) with \(t_N = t_T\). \(\mathcal{E}_{pk}\) (\(\mathcal{E}_{sk}\)) represents the pulse strength, \(f_{pk}(t)\) (\(f_{sk}(t)\)) is the profile function, \(\zeta_k\) is the center time  with \(\zeta_k = \left(k - \frac{1}{2}\right)\tau\), and \(\varphi_{pk}^0\) (\(\varphi_{sk}^0\)) refers to the initial phase of the \(k\)-th pump (Stokes) pulse.

The corresponding RWA Hamiltonian induced by the $k$-th pulse pair can be expressed as
\begin{eqnarray}
\begin{split}
\label{Eq:33}
&\hat{H}_{\text{R}}^{1\Lambda}(t_{k-1},t_k)=\\&-\frac{1}{2}\left( \Omega _{pk}(t)e^{-i\varphi _{pk}}\left\vert1\right\rangle \left\langle 1'\right\vert +\Omega _{sk}(t)e^{-i\varphi_{sk}}\left\vert 2\right\rangle \left\langle 1'\right\vert +\text{H.c.}\right)\text{,}
\end{split}
\end{eqnarray}
with the Rabi frequencies \(\Omega_{pk}(t) = \mu_{11'}\mathcal{E}_{pk} f_{pk}(t)\) and \(\Omega_{sk}(t) = \textbf{}\mu_{21'}\mathcal{E}_{sk} f_{sk}(t)\), and the phases \(\varphi_{pk} = \omega_p \zeta_k - \varphi _{pk}^0\) and \(\varphi_{sk} = \omega_s \zeta_k - \varphi _{sk}^0\).
The corresponding eigenstates of the Hamiltonian in Eq.~(\ref{Eq:33}) can be given by \(\left\vert \psi _{\pm}(\theta_k \left( t\right) )\right\rangle\) with eigenenergies \(E_{\pm k} (t)= \pm\frac{1}{2}\sqrt{\Omega_{pk}^{2} (t) + \Omega_{sk}^{2}(t) }\), and
\begin{eqnarray}\label{firstRaman}
\begin{split}
\left\vert \psi _{0}(\theta_k \left( t\right) )\right\rangle 
&=\cos \frac{\theta_{k}(t)}{2}\left\vert 1\right\rangle +e^{-i\varphi
_{k}+i\pi}\sin \frac{\theta_{k}(t)}{2}\left\vert
2\right\rangle,
\end{split}
\end{eqnarray}
with eigenenergie \(E_{0}(t)=0\), the relative phase is \(\varphi_{k} = \varphi_{sk} - \varphi_{pk}
\), and the mixing angle \(\theta_{k} (t)\) defined by \(\tan [\theta_{k} (t)/2]= {\Omega_{pk} (t)}/{\Omega_{sk} (t)}\). 
By defining the effective Rabi frequency $\Omega_{0k}(t) = \sqrt{\Omega_{pk}^2(t) +\Omega_{sk}^2(t)} $ that satisfies the relation  \(\Omega_{0k} = 2\sqrt{-E_{-k}E_{+k}}\), we can parameterize the corresponding Rabi frequencies as \(\Omega_{pk}(t) = \Omega_{0k}(t) \sin(\theta_{k}/2)\) and \(\Omega_{sk}(t) = \Omega_{0k}(t) \cos(\theta_{k}/2)\).
\\ \indent
We now derive the parameter conditions required to generate pulse sequences for the pulsed jump protocol. According to Eq.~(\ref{Eq:5}), nonadiabatic transitions encompass geometric and dynamical components. As indicated in Eq.~(\ref{Eq:7}), dynamical phase factors \(Y_{f,h}\) play a dominant role in minimizing nonadiabatic effects compared to the energy gaps \(E_{fh}\), which are not explicitly present in the expression. To fulfill the condition \(\epsilon \rightarrow 0\), the dynamical phase factors \(Y_{f,h}(\theta)\) at various parameter points \(\theta_{k}\) should add destructively along the evolution path. For \( g_{f,h} \neq 0 \) \((f = 0, h = \pm)\), we can modify 
\( Y_{f,h}(\theta) = \prod_{k=1}^{N} Y_{f,h}(\theta_k) \) 
to induce a \(\pi\)-phase shift at the parameter point \(\theta_k\), corresponding to the time  interval \( t \in \left( t_{k-1}, t_k \right) \), as depicted in Fig.~\ref{fig:0}(f). To ensure this, the dynamical phase factors should satisfy the condition
\begin{eqnarray}
  \label{Eq:dyp}
Y_{0,\pm}(\theta_k) = e^{i \left[ \alpha_\pm(\theta_k) - \alpha_0(\theta_k) \right]}=e^{\mp i(2s_0+1)\pi}=e^{\mp i\pi} \ (s_0\in\mathbb{N}_0), 
\end{eqnarray}
with the dynamical phases satisfy the condition $\alpha_\pm(\theta_k) = \alpha_\pm(t_{k-1},t_k) = \mp(2s_0+1)\pi$.
When \( g_{f,h} = 0 \) \((f,h=\pm)\) and the phase condition \(\dot{\varphi}_J (t)= 0\) is satisfied with \({\varphi}_J (t)= \sum_{k=1}^N{\varphi_k(t)}\), meaning the relative phase $\varphi_J(t)$ is a constant, \(Y_{f,h}(\theta)\) can be chosen in arbitrary functional forms. These discussions lead to $\hat{U}_{\text{Dia}}(\Theta_k) = P \exp[{i\int_{\theta_{0}}^{\Theta_{k}}W(\theta^{\prime})]d\theta^{\prime}}=\mathbb{I}$ with \(\Theta_{k}=2\sum_{v=1}^{k} (-1)^{v+k} \theta_{v}\). The evolution $\hat{U}$ after each pump-Stokes pulse pair realizes the desired `dark' state path described in Eq.~(\ref{Eq:10}) at $\Theta_{k}$ in the parameter space. For calculations, we assume that the $\theta_{k}$ points are equally spaced, following the expression 
\begin{eqnarray}\theta_{k} = \frac{\theta_{T}-\theta_{0}}{2N} \left( 2k-1 \right) + \theta_{0},
\label{Eq:theta}
\end{eqnarray}
with \(\theta_0 = 0\), and the $\Theta_{k}$ points are given by $\Theta_{k} = \frac{k\theta_{T}}{N}$ and $\Theta_N=\theta_T=\theta_{N+1}$.\\ \indent   
To satisfy the phase change requirements for $Y_{f,h}(\theta)$ 
 given in Eq.~(\ref{Eq:dyp}), we know that the eigenvalues are given by
$\int_{t_{k-1}}^{t_{k}} E_{-k}(t') \ dt' = -{(2s_0+1) \pi}$
and $\int_{t_{k-1}}^{t_{k}} E_{+k}(t') \ dt' = {(2s_0+1) \pi}$. 
It indicates that the effective subpulse area needs to meet a specific condition
\begin{eqnarray}
\begin{aligned}
\label{Eq:11}
\mathcal{A}_{0k}=\mathcal{A}_{0}(t_{k-1},t_k) &=\sqrt{\mathcal{A}^2_{pk}+\mathcal{A}^2_{sk}} \\ \
&=\int_{t_{k-1}}^{t_{k}} \Omega_{0k}(t') dt'=2(2s_0+1)\pi,
\end{aligned}
\end{eqnarray}
with the subpule areas $\mathcal{A}_{pk}=\int_{t_{k-1}}^{t_k}\Omega_{pk}(t)dt=2(2s_0+1)\pi\sin({\theta_k}/2)$ and $\mathcal{A}_{sk}=\int_{t_{k-1}}^{t_{k}}\Omega_{sk}(t)dt=2(2s_0+1)\pi\cos({\theta_k/2})$. This involves an effective pulse area $\mathcal{A}_{0k}=\sqrt{\mathcal{A}_{pk}^2+\mathcal{A}_{sk}^2}=2\pi$ when $s_0=0$, which represents the minimal pulse area required to achieve complete population transfer in a three-level $\Lambda$ system. \\ \indent
For simplicity,  we  consider each pump-Stokes pair with a Gaussian profile  
\begin{equation}
  \label{Eq:12}
 f_{p(s)k}(t)=f_{p(s)k0} e^{-\frac{(t - \zeta_k)^2}{2\tau_J^2}},
\end{equation}
where the pulse duration $\tau_J$ and the center time $\zeta_k$ are selected to ensure that there is no overlap between the pulse pairs $(\tau\ge6\sqrt{2}\tau_J)$, as illustrated in Fig.~\ref{fig:0}(c).
Furthermore, the pulse strengths $\mathcal{E}_{p(s)k}$ and the maximum effective Rabi frequency $\Omega_{0k}$ can be calculated to satisfy the effective pulse area in Eq.~(\ref{Eq:11}).\\ \indent
Since the relative phase depends on the pulse center time \(\zeta_k\), we need to adjust the initial phase of each pulse according to the determined \(\zeta_k\) to ensure the relative phase remains constant over time. We can simplify the analysis by employing a phase-locking operation and assuming \(\varphi_{p(s)k}^0 = 0\) (see Appendix~\ref{AppendixB}). With this assumption, the Hamiltonian in Eq.~(\ref{Eq:33}) becomes phase-independent, and in the following discussion, we implicitly apply this operation.

To further investigate the underlying dynamics of the evolution described, we analyze the contributions of the \(N\) pairs of composite pulse sequences. The unitary time-evolution operator that characterizes the dynamics induced by these pulse sequences reads
 \(\hat{U}\left(\theta_T\right) = \hat{U}_{\delta}\left(\theta_N\right) \hat{U}_{\delta}\left(\theta_{N-1}\right) \ldots \hat{U}_{\delta}\left(\theta_1\right)\), with \(\hat{U}_{\delta}\left(\theta_{k}\right)\) corresponding to \(\hat{U}_{\delta}\left(t\right)\) during the time interval \(t \in (t_{k-1}, t_k]\).
We can concisely describe the propagator \(\hat{U}_{\delta}(\theta_k)\) in the basis \(\left\{|1\rangle, |2\rangle, |1'\rangle\right\}\) as
\begin{widetext}
\begin{equation}
\begin{aligned}
  \label{Eq:17}
 \hat{U}_{\delta}[\theta_k(t)] &= \sum_{f=0,\pm}e^{-i\int_{t_{k-1}}^{t}E_{fk}(t')dt'}|\psi_{f}(\theta_{k})\rangle\langle\psi_{f}(\theta_{k})| \\
  &=\left( 
\begin{array}{ccc}
\cos ^{2}\left( \theta _{k}/2\right) +\sin ^{2}\left( \theta
_{k}/2\right) z_{k}(t) & \frac{1}{2}\sin \left( \theta
_{k}\right) \left[ z_{k}(t)-1\right]  & \sin \left( \theta _{k}/2\right) y_{k}(t)\\ 
\frac{1}{2}\sin \left( \theta _{k}\right) \left[z_{k}(t)-1\right]  & \sin ^{2}\left( \theta _{k}/2\right)
+z_{k}(t)\cos ^{2}\left( \theta _{k}/2\right)  & \cos \left(
\theta _{k}/2\right)y_{k}(t) \\ 
\sin \left( \theta _{k}/2\right) y_{k}^{\ast }(t) & \cos \left(
\theta _{k}/2\right)y_{k}^{\ast }(t) & z_{k}(t)
\end{array}%
\right) ,
\end{aligned}
\end{equation}
\end{widetext}
with the time-dependent Cayley-Klein parameters $z_{k}(t)=\cos {
[\mathcal{A}_{0k}(t)/2]}$ and $y_{k}(t)=-i\sin[\mathcal{A}_{0k}(t)/2]. $
The relevant parameters for achieving the desired evolution are determined and substituted through the pulsed jump protocol, ensuring that \(\hat{U}\left(\theta_T\right)\) corresponds to the desired unitary operator \(\hat{U}_{\text{Adia}}(\theta_T)\) within a finite time \(t_T = N\tau\). This approach enables the achievement of a unitary operator at the final moment,
\begin{eqnarray}
\begin{aligned}
  \label{Eq:18}
&\hat{U}\left[\theta_T(t_T)\right]
=\\&\left( 
\begin{array}{ccc}
\cos \left( \Theta _{N}/2\right) & \left( -1\right) ^{N}\sin \left( \Theta
_{N}/2\right)& 0 \\ 
-\sin \left( \Theta _{N}/2\right) & \left( -1\right) ^{N}\cos
\left( \Theta _{N}/2\right) & 0 \\ 
0 & 0 & \left( -1\right) ^{N}
\end{array}
\right).
\end{aligned}
\end{eqnarray}
Based on the discussion in Eq.~(\ref{Eq:18}), when considering only the resonant three-level system, it is evident that after applying \(N\) pulses designed according to the digital jump protocol, the wave function at the final time can be expressed as \( \left\vert \Psi(t_T) \right\rangle = \cos(\Theta_{N}/2) \left\vert 1 \right\rangle+\exp{(i\pi)}\sin(\Theta_{N}/2)\left\vert 2 \right\rangle \) with $|\Psi(t_0)\rangle=|1\rangle$, indicating that the system successfully reaches the `dark' state at the $t_T$ through the implementation of the pulsed jump protocol. 
\subsection{Analysis of the pulsed jump protocol in a double $\Lambda$ system}
We now analyze a four-level model by introducing an excited state $|2'\rangle$ closely above the state $|1'\rangle$. When we apply the pulse sequence, initially designed for the three-level system, to this expanded four-level system, the corresponding Hamiltonian under the RWA  reads
\begin{equation}
\begin{split}
  \label{Eq:21}
&\hat{H}_{\text{RWA}}(t_{k-1},t_k)=\\&-\frac{1}{2}\left(\begin{array}{cccc}0&0&\Omega_{pk}(t)&\Omega_{pk}'(t)\\0&-2\Delta_2&\Omega_{sk}(t)&\Omega_{sk}'(t)\\\Omega_{pk}(t)&\Omega_{sk}(t)&0&0\\\Omega_{pk}'(t)&\Omega_{sk}'(t)&0&-2\Delta_1\end{array}\right)
\end{split},
\end{equation}
where the Rabi frequencies are \(\Omega^{\prime}_{pk}(t) = \mu_{12'}\mathcal{E}_{pk} f_{pk}(t)\) and \(\Omega^{\prime}_{sk}(t) = \mu_{22'}\mathcal{E}_{sk} f_{sk}(t)\), the one-photon detuning is \(\Delta_1 = \omega_{2'1} - \omega_p=\omega_{2'2}-\omega_s\), and the two-photon detuning is \(\Delta_2 = \omega_{21} + \omega_s - \omega_p\).
\\ \indent
To analyze how the state $|2'\rangle$ affect the target state $|2\rangle$, we apply the pump-Stokes pulse sequence in Eq.~(\ref{pulse}) to the second  $\Lambda$ system consisting of three states $|1\rangle$, $|2\rangle$  and $|2'\rangle$. When the two-photon detuning is zero, the corresponding RWA Hamiltonian is given by
\begin{eqnarray} \label{2Lambda}
\begin{split}
&\hat{H}_{\text{R}}^{2\Lambda}(t_{k-1},t_k)=\\&-\frac{1}{2}\left( \Omega^{\prime}_{pk}(t)\left\vert1\right\rangle \left\langle 2'\right\vert +\Omega^{\prime}_{sk}(t)\left\vert 2\right\rangle \left\langle 2'\right\vert+\text{H.c.}\right)+\Delta_1|2'\rangle\langle 2'|\text{.}
\end{split}
\end{eqnarray}
The orthonormal eigenstates of the Hamiltonian in Eq.~(\ref{2Lambda}) are given by 
\begin{eqnarray} \label{secondRaman}
\begin{split}
\left\vert \psi^{\prime} _{-}\left( \theta_k^{\prime}(t)\right) \right\rangle =&\sin \frac{\theta_{k}^{\prime}(t)}{2}\cos \xi(t) \left\vert 1\right\rangle +\sin \xi(t)
\left\vert 2^{\prime }\right\rangle \\ &+\cos
 \frac{\theta_{k}^{\prime}(t)}{2}\cos \xi(t) \left\vert 2\right\rangle\text{,} \\
\left\vert \psi^{\prime} _{+}(\theta_k^{\prime} \left( t\right) )\right\rangle 
=&\sin\frac{\theta_{k}^{\prime}(t)}{2}\sin \xi(t)
\left\vert 1\right\rangle -\cos \xi(t)\left\vert 2^{\prime }\right\rangle \\&+\cos \frac{\theta_{k}^{\prime}(t)}{2}\sin \xi(t)
\left\vert 2\right\rangle  \text{,} \\
\left\vert \psi^{\prime} _{0}(\theta_k ^{\prime}\left( t\right) )\right\rangle 
=&\cos \frac{\theta_{k}^{\prime}(t)}{2} \left\vert 1\right\rangle +e^{i\pi}\sin \frac{\theta_{k}^{\prime}(t)}{2}\left\vert
2\right\rangle ,
\end{split}
\end{eqnarray}
with eigenenergies $E^{\prime}_{\pm k}(t)= \frac{1}{2}\left[ \Delta_1 \pm \sqrt{\Delta_1^{2} + \Omega_{pk}^{'2} (t) + \Omega_{sk}^{'2}(t) }\right]$ and \(E^{\prime}_{0}(t)= 0\), where the mixing angle \(\theta^{\prime}_{k} (t)\) is defined as \(\tan [\theta^{\prime}_{k} (t)/2]= {\Omega^{\prime}_{pk} (t)}/{\Omega^{\prime}_{sk} (t)}\) and the second mixing angle \(\xi(t)\) is determined by \(\tan 2\xi  (t)= {\Omega^{\prime}_{0k} (t)}/{\Delta_1}\) with the effective Rabi frequency $\Omega^{\prime}_{0k} (t)=\sqrt{\left(\Omega_{pk}^{\prime} (t)\right)^2 + \left(\Omega_{sk}^{\prime}(t)\right)^2 }$.\\ \indent 
The presence of \(|2'\rangle\) introduces an additional Raman transition path that impacts the population of the \(|2\rangle\) state. As a result, the pulses designed based on the resonant three-level $\Lambda$ model cannot guide the system along the adiabatic paths by the jump protocol. By analyzing Eqs.~(\ref{firstRaman}) and (\ref{secondRaman}), the two Raman transitions may take place deconstructive interference effects by controlling their amplitudes and phases. Hence, it is still feasible to achieve jump transitions by adjusting key parameters of the pump and Stokes pulse pairs. 
\section{RESULTS AND DISCUSSION}\label{III}
\begin{figure*}[ht]
  \centering
  \includegraphics[width=0.7\textwidth]{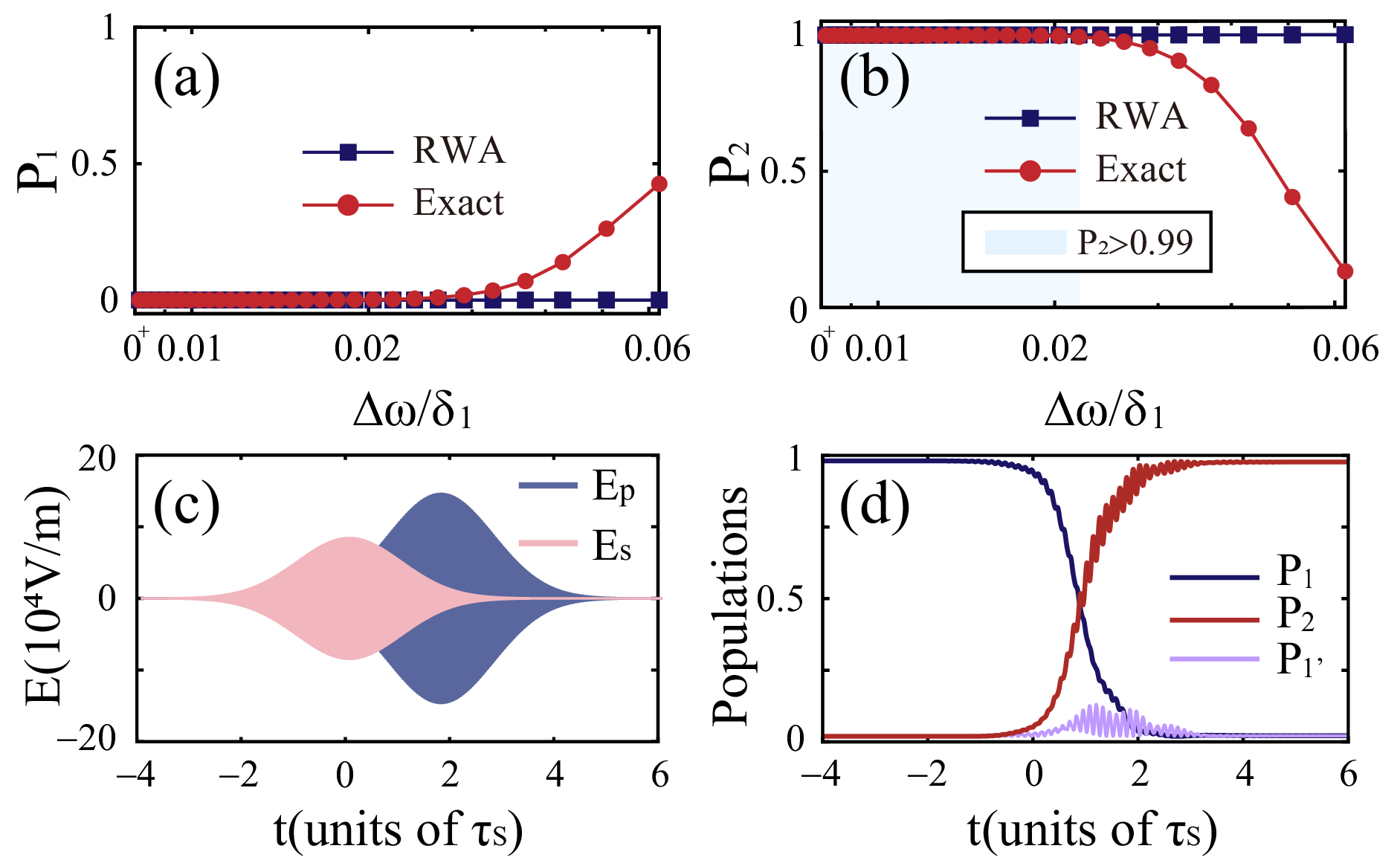}
  \caption{Quantum state  transfer in a single \(\Lambda\)  three-level system using STIRAP. Final populations in states $|1\rangle$  (a) and $|2\rangle$ (b) versus the bandwidths of the pulses for STIRAP. (c) The pulse sequence of the pump and Stokes driving fields, \( \mathcal{E}_p(t) \) and \( \mathcal{E}_s(t) \) with $\Delta\omega=0.02\delta_1$, and  (d) the corresponding time-dependent populations in three states. The results calculated using the RWA (blue line with squares) are compared with exact numerical simulations (red line with circles) in  (a) and (b).}
  \label{fig2}
\end{figure*}
We perform simulations to examine our methods in the prototype four-level double-$\Lambda$ system of ultracold $^{87}\mathrm{Rb}$ atoms, which is well
studied theoretically and experimentally \cite{WangDongsheng_PRA_2011,Namazi_PRAppl_2017,Saesun_OE_2018,RRS_PRL_(2019),Sagona_PRL_2020}. 
The system comprises the hyperfine levels $F=1$ and $2$ of the ground electronic state $5^{2}S_{1/2}$, with an energy split of $\delta_{1} = 6.8347\ \text{GHz}$, and the hyperfine levels $F^{\prime}=1'$ and $2'$ of the excited electronic state $5^{2}P_{1/2}$, with an energy split of $\delta_{2} = 0.8145\ \text{GHz}$. By applying linearly polarized pulses to the system, the dipole-allowed transitions between the magnetic sublevels $m_{F}$ and $m_{F^{\prime}}$ of the hyperfine levels $F$ and $F^{\prime}$ follow the selection rule of $\Delta m_{F}=0$. Considering the presence of hyperfine coupling, the eigenstates of the Hamiltonian are represented by the basis $\left\vert nlsjIFm_{F}\right\rangle$. Within this framework, \(n\) represents the principal quantum number, \(l\) the orbital angular momentum quantum number, \(s\) the spin quantum number, \(j\) the total electronic angular momentum quantum number with \(j = l + s\), \(I\) the nuclear spin quantum number, \(F\) the total angular momentum quantum number with \(F = j + I\), and \(m_{F}\) the magnetic quantum number describing the projection of \(F\) along a specified axis.
\\ \indent 
For convenience, the four hyperfine levels are described by the states \( |1\rangle \), \( |2 \rangle \), \( |1'\rangle \), and \( |2'\rangle \), corresponding to the energies \( E_{1} \), \( E_{2} \), \( E_{1'} \), and \( E_{2'} \), respectively. The time-dependent Hamiltonian of the system in the presence of the control fields $\mathcal{E}(t)=\sum_{k=1}^N\left[\mathcal{E}_{pk}(t)+\mathcal{E}_{sk}(t)\right]$ is given by \(\hat{H}(t) = \hat{H}_{0} -\hat{\mu} \mathcal{E}(t)\), where the field-free Hamiltonian is \(\hat{H}_{0} = \sum_{F=1,2} E_{F} |F \rangle \langle F| + \sum_{F'=1',2'} E_{F'} |F' \rangle \langle F'|\), and the matrix elements of the dipole operator \(\hat{\mu}\) are \(\mu_{FF'} = \langle F|\hat{\mu}|F'\rangle\).
The evolution of the system from the initial time \( t_{0} \) to a given time \( t \) can be described by using the unitary operator \( \hat{U}(t) \) with \( \hat{U}(t_0) = \mathbb{I} \), which satisfies the Schrödinger equation \(\left( \hbar = 1 \right)\)
\begin{equation}
  \label{Eq:0}
i\frac{d}{dt}{\hat{U}}(t)=\hat{H}(t)\hat{U}(t).
\end{equation}
As direct transitions between states $|1\rangle$ and $|2\rangle$ are not allowed, two Raman transition pathways from state $\left\vert 1\right\rangle$ to state $\left\vert 2\right\rangle$ via the intermediate states $\left\vert 1' \right\rangle$ and $\left\vert 2' \right\rangle$ are involved. The elements of the transition dipole operator $\hat{\mu}$ between states satisfy the relations $\mu_{12'} = -\sqrt{3} \mu_{11'} = \sqrt{1/4} \mu_{J}$ and $\mu_{21'} = \sqrt{3} \mu_{22'} = \sqrt{1/4} \mu_{J}$, where $\mu_{J}$ represents the transition dipole matrix element of the electronic transition from the ground electronic state $5^{2}S_{1/2}$ to the excited electronic state $5^{2}P_{1/2}$ (see Appendix \ref{AppendixA}). By considering the system initially in the ground state $|1\rangle$, the time-dependent wave function of the system can be obtained by $|\Psi(t)\rangle=\hat{U}(t)|\Psi(t_0)\rangle$ with $|\Psi(t_0)\rangle=|1\rangle$. 
\subsection{Simulations in a three-level $\Lambda$ system}
\begin{figure*}[ht]
  \centering
  \includegraphics[width=0.7\textwidth]{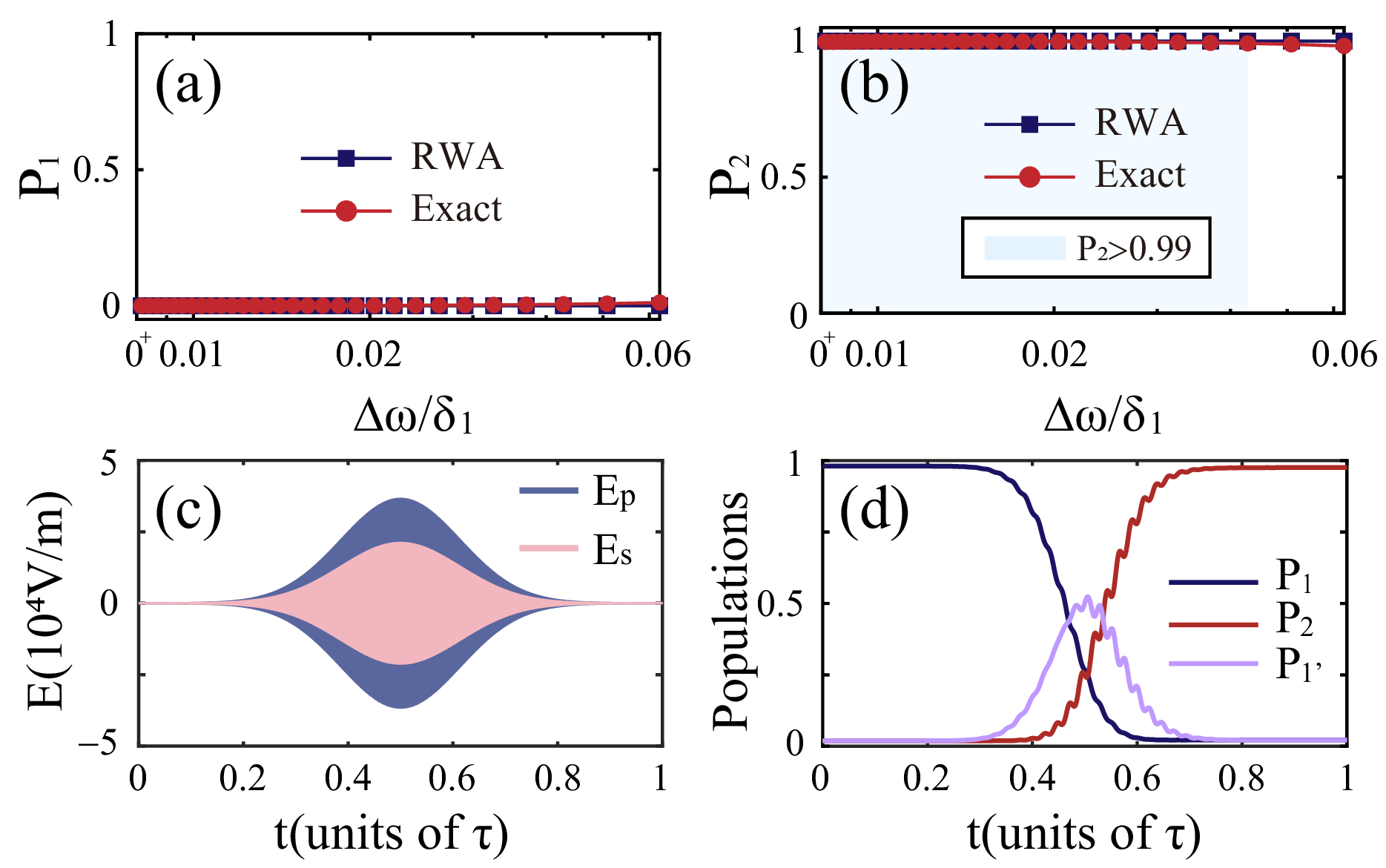}
  \caption{Quantum state transfer in a single \(\Lambda\)  three-level system using the pulsed jump protocol with $N=1$. Final populations in states $|1\rangle$  (a) and $|2\rangle$ (b) versus the bandwidths of the pulses for the pulsed jump protocol. (c) The pulse sequence of the pump and Stokes driving fields, \( \mathcal{E}_{p}(t) \) and \( \mathcal{E}_s(t) \) with $\Delta\omega=\delta_1/30$, and (d) the corresponding time-dependent populations in three states. The results in parts (a) and (b), calculated using the RWA (blue line with squares), are compared with exact numerical simulations (red line with circles).}
  \label{fig3}
\end{figure*}
We first  explore how to transfer the population from the ground state $|1\rangle$ to the target state $|2\rangle$ using STIRAP in a three-level system consisting of states $|1\rangle$, $|2\rangle$, and $|1'\rangle$, while excluding state $|2'\rangle$, as illustrated in Fig.~\ref{fig:0}(a). To explore the validity of the RWA model and its dependence on the bandwidth \(\Delta \omega\) of the two pulses, we consider the pump and Stokes pulses in the following forms  
\begin{equation}\label{psp}
\begin{aligned}
&\mathcal{E}_p(t)=\sqrt{\frac{1}{2\pi}}\frac{\mathcal{A}_p\Delta \omega}{\mu_{11'}}e^{-\frac{1}{2}(t-t_{p})^{2}\Delta \omega^{2}}\cos{(\omega_{p}t)},\\
&\mathcal{E}_s(t)=\sqrt{\frac{1}{2\pi}}\frac{\mathcal{A}_s\Delta \omega}{\mu_{21'}}e^{-\frac{1}{2}(t-t_{s})^{2}\Delta \omega^{2}}\cos{(\omega_{s}t)},\\
\end{aligned}
\end{equation}
with center times \(t_p\) and \(t_s \), and center frequencies \(\omega_p=\omega_{1'1}\) and \(\omega_s=\omega_{2'1}\). The pulse duration \(\tau_S\) is given by \(\tau_S = 1/\Delta\omega\). The pulse areas in Eq.~(\ref{psp}) are defined as \(\mathcal{A}_p=\int_{t_0}^{t_T}\Omega_{p}(t)dt\) and \(\mathcal{A}_s=\int_{t_0}^{t_T}\Omega_s(t)dt\), with \(t_0=-4\tau_S\) and \(t_T=6\tau_S\). Note that this definition of the electric strengths in Eq.~(\ref{psp}) ensures that their pulse areas remain constants of  $\mathcal{A}_p$ and $\mathcal{A}_s$ as the pulse duration $\tau_S$ changes.\\ \indent
In the STIRAP technique, the higher pulse areas of the pump and Stokes pulses lead to the lower maximum transient population in the excited intermediate state. To illustrate, we consider the two pulse areas of \(\mathcal{A}_p=\mathcal{A}_s=6\sqrt{2}\pi\), resulting in an effective pulse area of \(\mathcal{A}_0=\sqrt{\mathcal{A}_p^2+\mathcal{A}_s^2}=12\pi\), which are significantly larger than the minimal pulse area of \(2\pi\) needed to achieve complete population transfer in a three-level system. The center times of the two pulses with ‘counterintuitive’ time delays are fixed at  \(t_s=0\) and \(t_p=1.76\tau_S\). Under these parameter conditions, we can achieve high efficiency of population transfer in the STIRAP system model. To calculate the time-dependent wave function \(\lvert \Psi^{\text{RWA}}(t) \rangle\) of the three-level system under the RWA, we numerically solve the time-dependent Schr\"{o}dinger equation with the Hamiltonian \(\hat{H}_{\text{RWA}}(t)\) in Eq.~(\ref{Eq:8}), in which the pump and Stokes pulses are involved in the corresponding transition separately. The time-dependent wave function \(|\Psi(t)\rangle\) of the system is obtained by solving Eq.~(\ref{Eq:0}), in which the total control field is used by  \(\mathcal{E}(t)=\mathcal{E}_p(t)+\mathcal{E}_s(t)\). 

Figures \ref{fig2}(a) and (b) illustrate the dependence of the final populations in states \( |1\rangle \) and \( |2\rangle \) on the bandwidths of the pulses. The RWA calculated populations \( P_{F}^{\text{RWA}}(t_{T}) = \left\vert \left\langle F \mid \Psi^{\text{RWA}}(t_{T}) \right\rangle \right\vert^{2} \)  are compared with the exactly calculated populations \( P_{F}(t_{T}) = \left\vert \left\langle F \mid \Psi(t_{T}) \right\rangle \right\vert^{2} \).  
We observe that the exactly calculated populations using the total control fields $\mathcal{E}(t)$ agree with the RWA results in the narrow-bandwidth regime. As bandwidth increases, the exact results diverge significantly from the RWA, independent of the pulse bandwidths.  The fidelity of the state $|2\rangle$ can reach a value higher than 0.99 with a limit condition of $\Delta\omega<\delta_1/45$.  When the pulse bandwidths become broad, approaching the energy level spacing $\delta_1$ between states $|1\rangle$ and $|2\rangle$, the pump pulse that is used to excite levels $|1\rangle$ and $|1'\rangle$ inadvertently causes excitation between states $|2\rangle$ and $|1'\rangle$. Similarly, the Stokes pulse, exclusively employed for exciting states $|2\rangle$ and $|1'\rangle$, can also lead to the excitation of states $|1\rangle$ and $|1'\rangle$. As a result, the pulses in the broad bandwidth regime break the conditions for using the RWA method. Figures \ref{fig2}(c) and (d) show the time-dependent electric fields of the pump and Stokes pulses at a bandwidth of $\Delta \omega=\delta_1/50$, along with the corresponding population evolutions of three states. The pulses successfully transfer the system from the ground state $|1\rangle$ to the target state $|2\rangle$ with a high fidelity of over 0.995. The slight transient population in the excited intermediate state can be further suppressed by increasing the pulse areas while decreasing the bandwidths.

We now examine the population transfer of a three-level system by using the pulsed jump protocol described in Section \ref{D}.  To find how the model depends on the bandwidth of each subpulse pair defined in Eqs.~(\ref{pulse}) and (\ref{Eq:12}), the electric field strengths are defined as $\mathcal{E}_{pk}=\sqrt{\frac{1}{2\pi}}\frac{\mathcal{A}_{pk}\Delta\omega}{\mu_{11'}}$ and $\mathcal{E}_{sk}=\sqrt{\frac{1}{2\pi}}\frac{\mathcal{A}_{sk}\Delta\omega}{\mu_{21'}}$ with $\Delta\omega=1/\tau_J$, where the values of the pulse areas $\mathcal{A}_{pk}$ and $\mathcal{A}_{sk}$ have been given by Eq.~(\ref{Eq:11}) with \(s_0 = 0\). To calculate the corresponding time-dependent wave function \(\lvert \Psi^{\text{RWA}} (t) \rangle\) of the three-level system under the RWA, we can numerically solve the time-dependent Schr\"{o}dinger equation with the Hamiltonian \(\hat{H}_{\text{RWA}}(t)=\sum_{k=1}^N\hat{H}_{R}(t_{k-1}, t_k)\) defined in Eq.~(\ref{Eq:33}). We solve  Eq.~(\ref{Eq:0}) to obtain time-dependent wave function \(|\Psi(t)\rangle\) of the system by using  the total control field $\mathcal{E}(t)=\sum_{k=1}^N\left[\mathcal{E}_{pk}+\mathcal{E}_{sk}\right]$ with $t_0=0$ and $t_T=N\tau=2.8\pi N\tau_J$. Note that the center times of the two pulses are applied simultaneously with \(\zeta_k=2.8\pi(k-1/2)\tau_J\). \\ \indent 
\begin{figure*}[ht]
  \centering
  \includegraphics[width=0.8\textwidth]{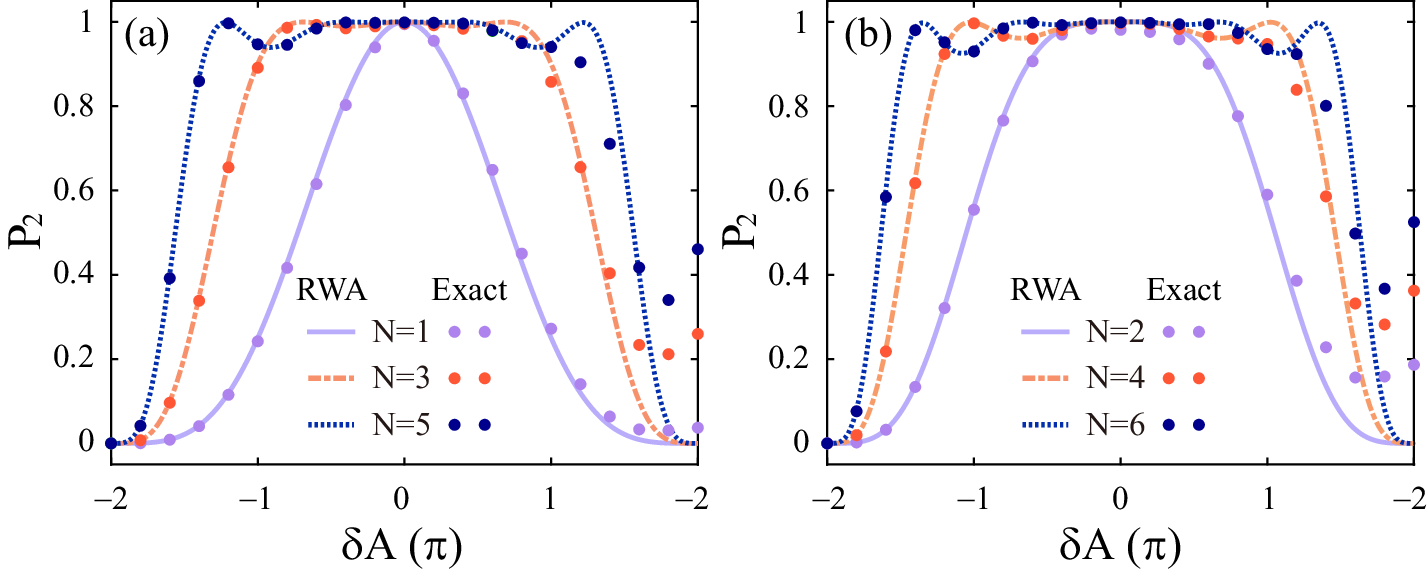}
  \caption{Robustness of the transfer efficiency as a function of the pulse area error $\delta \mathcal{A}$ for the jump sequence with (a) odd pulse pairs ($N = 1, 3, 5$) and (b) even pulse pairs ($N = 2, 4, 6$). The RWA (solid lines) results are compared with exact numerical simulations (dotted lines).}
  \label{fig4}
\end{figure*}
Figures \ref{fig3}(a) and (b) display the same comparison as those in  Figs.~\ref{fig2} (a) and (b) for the scenario of $N=1$.  The exactly calculated populations match well with the RWA results in the narrow-bandwidth regime but differ in the broad-bandwidth regime, where the conditions required for applying the RWA method are not satisfied. Note that the pulse bandwidths needed to achieve a high fidelity of $P_2(t_T)>0.99$ for the pulsed jump protocol can be broader than that of the STIRAP method with a limit condition of  $\Delta \omega < \delta_1/22$. Figure~\ref{fig3}(c)  shows the time-dependent electric fields of the two pulses at $\Delta \omega = \delta_1/30$, which leads to the same fidelity of over 0.995 in the state $|2\rangle$. As shown in Fig.~\ref{fig3}(d), which depicts the population evolution in the three states, the scheme successfully transfers a high-fidelity population to the target state at the end of the pulses. However, there is a noticeable transient population in the intermediate state during pulse overlap time. As a result, its robustness against control parameter fluctuations may be compromised. Increasing the number of pulse pairs is essential to demonstrate the advantage of the pulsed quantum jump protocol, thereby further increasing the total effective pulse area.\\ \indent
We now explore the robustness of the pulsed jump protocol by fixing the bandwidth at \(\Delta\omega = \delta_1/30\) as an illustrative case and increasing the number of subpulse pairs. Based on our theoretical analysis, the concept of the effective area \(\mathcal{A}_{0k}\) for each pulse pair is critical in understanding how fluctuations in control parameters affect control efficiency, which depends on several factors, including the electric field strengths \(\mathcal{E}_{pk}\) and \(\mathcal{E}_{sk}\), the pulse duration \(\tau_J\), the transition dipole moments \(\mu_{11'}\) and \(\mu_{21'}\), and the central frequencies \(\omega_{p}\) and \(\omega_s\). To see how its variation affects the control efficiency, we define the effective pulse area error for the pump and Stokes pulse pairs as \(\delta\mathcal{A} = \mathcal{A}_{0k} - 2\pi\).
Figures~\ref{fig4}(a) and (b) illustrate the population comparisons of \(P_{2}(t_{T})\) with \(P_{2}^{\text{RWA}}(t_{T})\) for the target state \(|2\rangle\) as a function of  \(\delta\mathcal{A}\), for both even and odd values of \(N\). It is observed that increasing the number of pulse pairs enhances the robustness of control efficiency. By comparing the exactly calculated populations with those obtained through RWA results, we observe that increasing the pulse area significantly beyond the optimal value of \(2\pi\) leads to notable discrepancies. This indicates that the pump and Stokes pulses cannot be distinctly separated when dealing with strong pulse pairs.\\ \indent 
\begin{figure}
  \centering
  \includegraphics[width=0.45\textwidth]{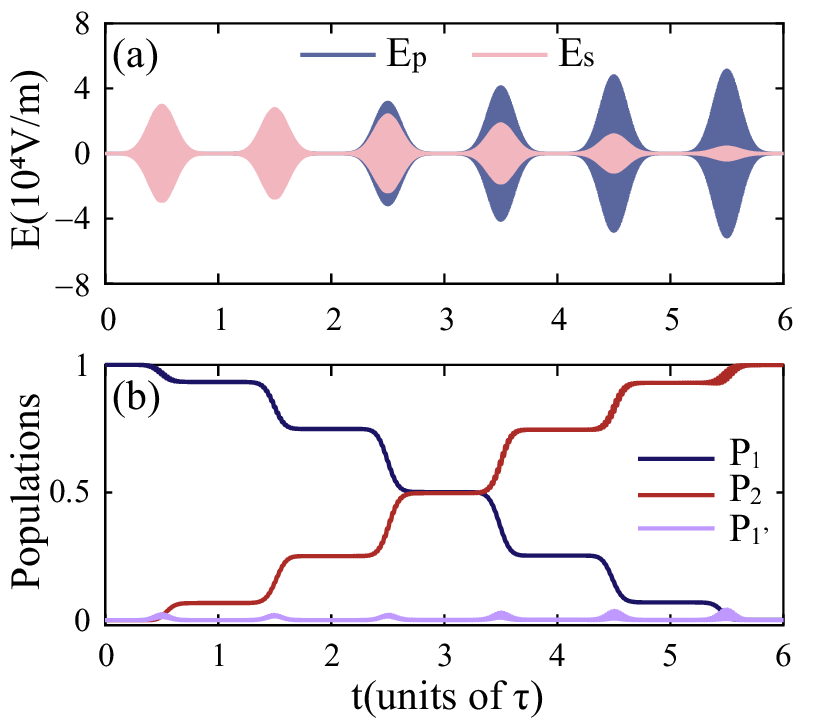}
  \caption{Population transfer in a resonant $\Lambda$ three-level system using the pulsed jump protocol with \( N = 6 \). (a) The pulse sequences of the pump and Stokes driving fields, \( \mathcal{E}_{p}(t) \) and \( \mathcal{E}_{s}(t) \). (b) The corresponding time-dependent populations. }
  \label{fig5}
\end{figure}
\begin{figure}[htbp]
  \centering
  \includegraphics[width=0.45\textwidth]{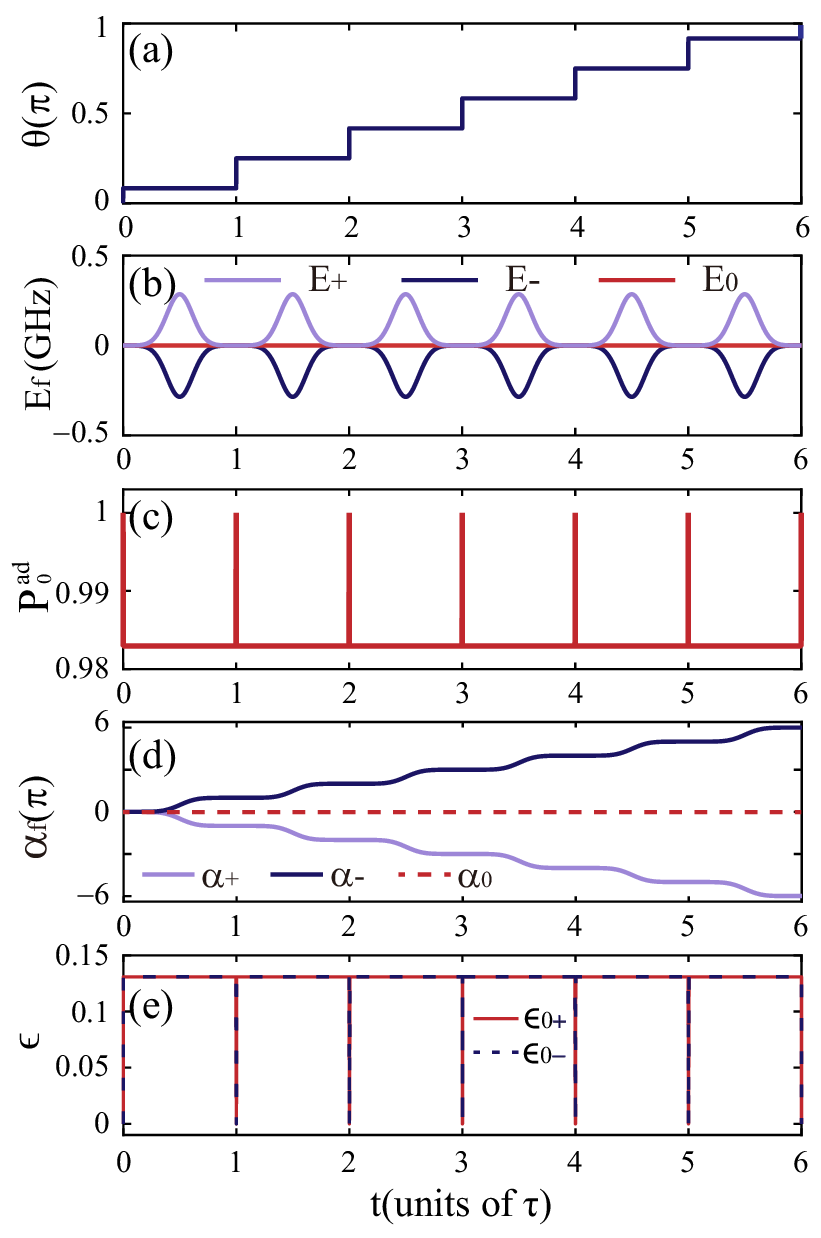}
  \caption{Intrinsic characterization of the pulse jump protocol with the adiabatic process. The time evolution of (a) the system's three eigenstates \( E_{\pm,0}(t) \), (b) mixing angle \( \theta(t) \), (c) adiabatic probability amplitude \( P^{\mathrm{ad}}_0(t) \), (d) dynamical phases \( \alpha_{\pm,0}(t) \), and (e) dynamic accumulation functions \( \epsilon_{0\pm}(t) \). }
  \label{fig6}
\end{figure}
To visualize how the system is transferred from the initial state \(|1\rangle\) to the target state \(|2\rangle\) via the pulsed jump approach, Figs.~\ref{fig5}(a) and (b) display the time-dependent pump and Stokes trains at \(N=6\), along with the corresponding population evolutions of the three states. We can observe that the involved six pulse pairs in Fig.~\ref{fig5}(a) are distinctly separated and do not overlap. As the number of pulse pairs increases, the maximum field strengths of the Stokes pulses gradually diminish while those of the pump pulses rise. From the overall interaction process perspective, the Stokes and pump pulses display a `counterintuitive' time delay despite having identical central times for each pulse pair. As shown in Fig.~\ref{fig5}(b), the system transitions from its initial state to the target state in a jump manner while effectively minimizing the occupancy of the intermediate state. 
\\ \indent 
To analyze the intrinsic characteristics of the pulsed jump protocol and its relationship with the adiabatic process, we examine its behavior in the adiabatic representation. Figure \ref{fig6}(a) illustrates how the parameter \(\theta\) varies stepwise during the pulsed jump protocol. Since the pump and Stokes pulse pairs are applied to the system simultaneously, the corresponding parameter point \(\theta_k\) remains constant within each period. This indicates that the pulsed jump protocol does not strictly satisfy the adiabatic condition, which requires that the rate of change of the system's parameter, \(\lambda(t)\), be much smaller than the energy gap \(E_{\pm} - E_0\), i.e., \(|\lambda(t)| \ll |E_{\pm}(t)|\).\\ \indent 
Figures \ref{fig6}(b) and (c) show the time-dependent evolution of the three adiabatic eigenvalues \(E_{\pm}(t) = \sum_{k}^{N} E_{\pm k}(t)\) of the Hamiltonian from Eq.~(\ref{Eq:33}), as well as the corresponding population \(P^{\text{ad}}_{0}(t) = \vert \langle \psi_0(t)|\Psi^{\text{RWA}}_{\text{ad}}(t) \rangle \vert^{2}\) in the adiabatic state \(|\psi_0\rangle\) with its eigenenergy \(E_0 = 0\). In these simulations, the adiabatic wavefunction is calculated using \(|\Psi^{\text{RWA}}_{\text{ad}}(t)\rangle = \hat{U}_{a}^\dagger |\Psi^{\text{RWA}}(t)\rangle\), which involves a unitary transformation matrix \(\hat{U}_{a}\) \cite{Baksic_PRL_2016,Chathanathil_PRA_2023,Chathanathil_2023_QST}. We observe that the evolution process deviates from the state \(|\psi_0(t)\rangle\) within the time interval \((t_{k-1}, t_k)\), indicating the presence of non-adiabatic contributions. This is consistent with a small population in the intermediate state \(|1'\rangle\), as shown in Fig.~\ref{fig5}(b). Due to the constancy of \(\theta_k\) within each period, the population \(P^{\text{ad}}_{0}(t)\) in the state \(|\psi_0\rangle\) remains unchanged. At the end of the pulse, the non-adiabatic contributions interfere destructively,  averaging to zero. This allows the evolution to align with the desired path, such that \(\hat{U}(t_k) \approx \hat{U}_{\text{Adia}}(t_k)\). 
\\ \indent 
We now examine whether the control scheme meets the necessary and sufficient adiabatic conditions. Figure~\ref{fig6}(d) shows the time-evolution of the relative dynamical phases \(\alpha_{\pm,0}\). It exhibits a step-like variation concerning the mixing angle \(\theta\) and accumulates a phase of \(\pi\) after each pair of pulses. This behavior indicates that the dynamic phase factor \(Y_{0,\pm}\) satisfies the conditions specified in Eq.~(\ref{Eq:dyp}) and demonstrates periodic oscillations. Figure \ref{fig6}(e) depicts the time-dependent evolution of the necessary and sufficient condition \(\epsilon_{0\pm}(t) = \left| \int_{t_0}^{t} Y_{0,\pm}(t') \lambda \, dt' \right|\). We find that \(\epsilon_{0 \pm}\) attains non-zero values within each time interval and approaches zero at the time points \(t_k\), which correspond to the path points \(\Theta_k\). Consequently, the evolution path of the pulsed jump protocol is not entirely adiabatic because it mitigates nonadiabatic contributions through the dynamic modulation of the functional forms of the parameters. Furthermore, regardless of the number of pulse pairs employed, the desired target state can still be achieved at the points \(t_k\) (\(\Theta_k\)). Increasing the number of pulse pairs enhances the effective pulse area, resulting in more path points transitioning into the ideal adiabatic regime during the evolution.
\subsection{Simulations in a four-level double-$\Lambda$ system}
\begin{figure}[htbp]
  \centering
  \includegraphics[width=0.45\textwidth]{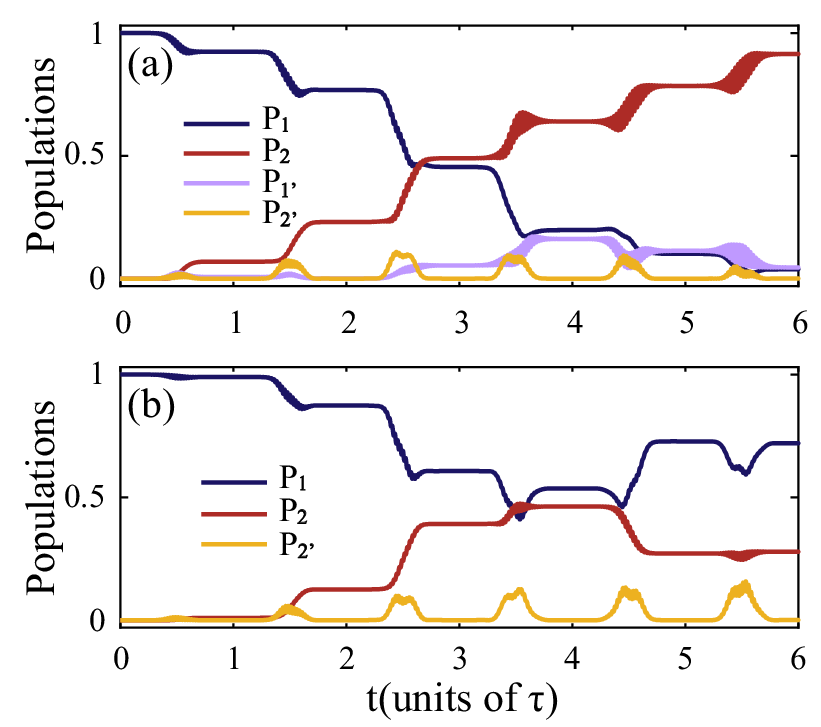}
  \caption{The time-dependent populations of the double $\Lambda$ four-level system (a) and the detuned single $\Lambda$ three-level system (b) using the same pulses as that in Fig.~\ref{fig5}. }
  \label{fig7}
\end{figure}
We now extend the pulsed jump protocol to the four-level double-$\Lambda$ system by incorporating the level \(|2'\rangle\), which is close to the level \(|1'\rangle\). According to our model, the energy difference \(\delta_{2} = 0.8145\ \text{GHz}\) between the hyperfine energy levels \( |1'\rangle \) and \( |2'\rangle \) is significantly smaller than \(\delta_{1} = 6.8347\ \text{GHz}\) between the hyperfine energy levels \( |1\rangle \) and \( |2\rangle \). This indicates that the pairs of jump pulses used for the resonant three-level \(\Lambda\) system may also interact with the level \( |2'\rangle \), opening the second three-level \(\Lambda\) transition path from \( |1\rangle \) to \( |2\rangle \) via \( |2'\rangle \).  
\begin{figure}
  \centering
  \includegraphics[width=0.45\textwidth]{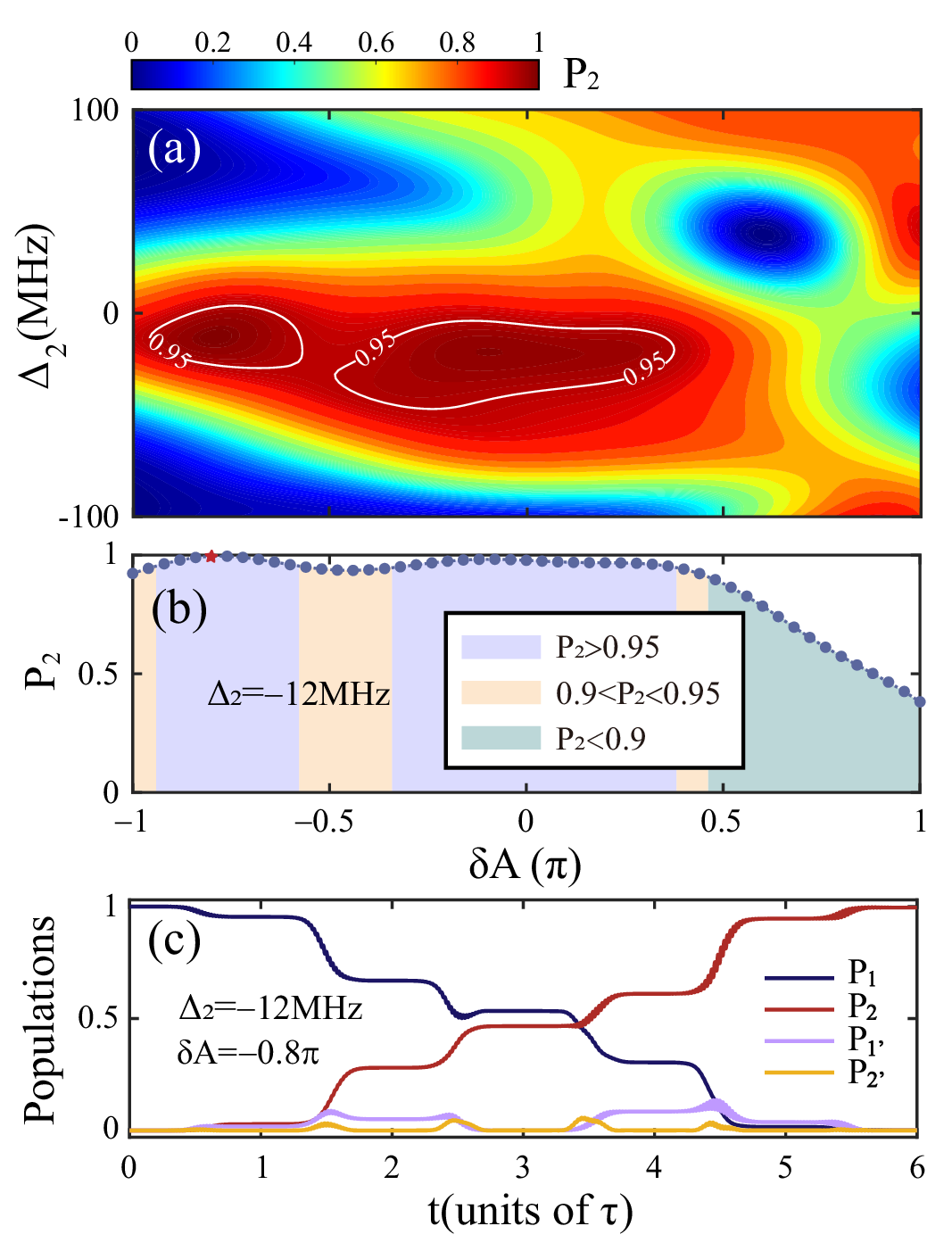}
  \caption{Quantum state transfer in a double \(\Lambda\) four-level system using the parameter-modulated pulsed jump protocol with \( N = 6 \). (a) Final population \( P_2(t_T) \) versus the pulse area variation \(\delta \mathcal{A}\) and (b) the two-photon detuning \(\Delta_2\), and the corresponding cut lines at  \(\Delta_2 = -12~\mathrm{MHz}\). (c) The corresponding time-dependent populations of four states driven by the pulses at \(\delta \mathcal{A} = -0.8\pi\) and \(\Delta_2 = -12~\mathrm{MHz}\).}
 \label{fig8}
\end{figure} 
Figure~\ref{fig7} illustrates the time-dependent populations of the four states \(|1\rangle\), \(|2\rangle\), \(|1'\rangle\), and \(|2'\rangle\), along with the three detuned states \(|1\rangle\), \(|2\rangle\), and \(|2'\rangle\) driven by the same pulse used in Fig.~\ref{fig5}. The optimized pulse pairs designed for the three-level system do not achieve high-fidelity population transfer in the four-level system, highlighting the impact of the neighboring level \(|2'\rangle\). During the pulse interaction, we observe that both intermediate states, \(|1'\rangle\) and \(|2'\rangle\), exhibit considerable transient populations. In contrast, the evolution of the population in state \(|2\rangle\) shows oscillatory fluctuations rather than a straightforward step-like increase. Furthermore, Fig.~\ref{fig7}(b) demonstrates that the neighboring state \(|2'\rangle\) significantly influences the mechanism of population transfer. However, this effect on the overall fidelity of the population transfer in the four-level system is relatively minor compared to the performance of the detuned three-level system, resulting in approximately an 8.6\% change. This raises a fundamental question about whether the negative effects of the neighboring state \(|2'\rangle\) on final efficiency can be mitigated through quantum interference effects,  where different transition paths interact, resulting in constructive or destructive interference patterns.  In our previous work \cite{RRS_PRL_(2019)}, we demonstrated a phenomenon of complete destructive interference under broad bandwidth excitation conditions. In this context, the double $\Lambda$ four-level system is reduced to a $V$-type three-level system, comprising the states $|1\rangle$, $|1'\rangle$, and $|2'\rangle$. It leads to the complete cancellation of population transfer to the target state $|2\rangle$.   \\ \indent 
 In the current four-level system, these interference effects may cancel out unwanted excitation impacts, thereby improving the efficiency of transferring the population to the target state $|2\rangle$. To enhance the quantum state transfer process, it is crucial to identify the optimal combination of parameters, such as detunings, relative phases, and time delays of the sub-pulse pairs. This work demonstrates that the pulse area and two-photon detuning are crucial parameters in the pulsed jump protocol. Figure~\ref{fig8}(a) shows the exactly calculated population of the target state as a function of variations in the pulse area \(\delta \mathcal{A}\) and two-photon detuning \(\Delta_2\). The other pulse parameters remain consistent with those presented in Fig.~\ref{fig5}. The results show that adjusting the pulse amplitude and introducing a two-photon red detuning can achieve a high population transfer in the four-level system. Figure \ref{fig8}(b) illustrates how \(P_{2}(t_{T})\) varies with \(\delta \mathcal{A}\) while keeping the two-photon detuning at \(\Delta_2 = -12\ \text{MHz}\). \(P_{2}(t_{T})\) remains largely unaffected by changes in amplitude, achieving over 95\% quantum state transfer across a broad range of pulse area variations, for which the population of the target state reaches a high value of 99.6\% at \(\delta\mathcal{A} = -0.8\pi\). This is also depicted in the corresponding time-dependent populations of the four states shown in Fig.~\ref{fig8}(c). While the state \(|2'\rangle\) participates in the population transfer process, its final effect is negligible due to quantum interference effects.
\section{CONCLUSION}\label{IV}
We developed a pulse-driven jump protocol to achieve all-optical Raman control over ultracold atomic hyperfine states. By establishing general conditions for adiabatic evolution within parameter space, we derived the necessary pulse area and phase conditions for the pulsed jump protocol, which is essential for quantum state transfer in a resonant single-$\Lambda$ three-level system utilizing an adiabatic evolution path along STIRAP. This approach was extended to a double-$\Lambda$ four-level system by incorporating a single-photon detuned $\Lambda$ three-level system by adding a neighboring intermediate state. By applying the derived conditions to ultracold $^{87}$Rb atoms, we clarified the bandwidth requirements for the STIRAP and pulsed jump protocol schemes derived from the rotating wave approximation. Our results demonstrated that high-fidelity and robust control over quantum state transfer can be achieved in a single-$\Lambda$ three-level system using both STIRAP and pulsed jump protocols. Moreover, we found that the destructive quantum interference effects between resonant and detuned Raman pathways in the double-$\Lambda$ four-level system can be alleviated by appropriately modulating the pulse area and two-photon detuning parameters within the pulsed jump protocol, leading to high-fidelity population transfer to the target state. This work proposes an effective method for achieving all-optical Raman control of quantum state transfer in ultracold atomic hyperfine states, offering significant theoretical and technical support for potential applications in quantum information processing and precision measurements. 

\section*{Acknowledgments}
This work was supported by the National Natural Science Foundation of China (NSFC) under Grants 12274470. The numerical simulation  was partially conducted using computing resources at the High Performance Computing Center of Central South University.

\renewcommand{\appendixname}{APPENDIX}
\appendix
\section{\MakeUppercase{PHASE-LOCKING OPERATIONS}}
\label{AppendixB}
To enhance the control over the phase in double Gaussian pulse sequences, we perform a frequency-domain analysis of the control pulse sequence \cite{Monmayrant_JOPB_2010,ShuChuanCun_PRA_2016,Wilson_PRAppl_2018}, 
$ {E}_{pk}(\omega) = A_{pk}(\omega) \exp{\left[i\phi_{pk}(\omega)\right]}$ and $ {E}_{sk}(\omega) = A_{sk}(\omega) \exp{[i\phi_{sk}(\omega)]}$, with $\phi_{p(s)k}(\omega)=\phi_{p(s)k}^0+(\omega-\omega_{p(s)})\phi_{p(s)k}^1+\frac12(\omega-\omega_{p(s)})^2\phi_{p(s)k}^2+\cdots $ ($k=1,2\ldots N$). We assume that the fixed phase $\phi _{pk}^0=0$ and $\phi _{sk}^0=0$, and consider the time-domain linear phase of $\phi_{pk}^1 = d\phi_{pk}/d\omega |_{\omega=\omega_{p}}=-\zeta_k$ and $\phi_{sk}^1 = d\phi_{sk}/d\omega |_{\omega=\omega_{s}}=-\zeta_k$, while neglecting higher-order terms. As a result, the complex spectral fields can be expressed as
 \begin{eqnarray}
 \begin{split}
  \label{Eq:13}
 {E} _{p(s)}\left( \omega \right) &=A _{p(s){1}}\left( \omega\right) e^{-i\left( \omega -\omega _{p(s)}\right)\zeta_1}+ A _{p(s){2}}\left( \omega \right)e^{-i\left( \omega -\omega _{p(s)}\right) \zeta_2}+\\&
 \ldots  A _{p(s){N}}\left( \omega \right) e^{-i\left(\omega -\omega _{p(s)}\right) \zeta_N} \text{,}
 \end{split}
\end{eqnarray}
where
\begin{eqnarray}
A _{p{k}}\left( \omega \right)  =\frac{ \mathcal{A} _{pk}}{{ 2\pi}\mu_{11'}}e^{{-\frac{
\left( \omega -\omega _{p}\right) ^{2}}{2\Delta \omega ^{2}}}}
\end{eqnarray} and 
\begin{eqnarray}
A _{s{k}}\left( \omega \right)  =\frac{ \mathcal{A} _{sk}}{{ 2\pi } \mu_{21'}}e^{{-\frac{
\left( \omega -\omega _{s}\right) ^{2}}{2\Delta \omega ^{2}}}},
\end{eqnarray} 
with the pulse bandwidth $\Delta \omega =1/\tau_J$. By performing inverse Fourier transforms $\mathcal{E}_{p(s)}(t)=\text{Re}[\int_{-\infty}^{\infty} {E} _{p(s)}\left( \omega \right) \exp{(i\omega t)} d\omega] $, we can obtain the time-dependent $k$-th control pulse pairs in Eq.~(\ref{psp}).
\section{\MakeUppercase{TRANSITION DIPOLE MOMENT CALCULATION}}
\label{AppendixA}
In this Appendix, we deduce in detail the analytical expression of the transition dipole matrix for the hyperfine levels of the \({}^{87}\mathrm{Rb}\) atom \cite{Sobelman2012Atomic,Rb_date_2024,Steck2024,RbBOOK(2025)}. Under the electric dipole approximation and in the presence of hyperfine coupling, the eigenstates of the Hamiltonian are described by the basis $\left\vert nlsjIFm_{F} \right\rangle$. The transition dipole moment $\mu_{FF'}$ from the hyperfine state $\left\vert nlsjIFm_{F} \right\rangle $ to the hyperfine state $\vert n^{\prime }l^{\prime }s^{\prime }j^{\prime }I^{\prime
}F^{\prime }m_{F} ^{\prime }\rangle$ can be calculated by 
\begin{eqnarray}
\begin{split}
\mu_{FF'}&= \langle nlsjIFm_{F}|\hat{r}|n^{\prime}l^{\prime}s^{\prime}j^{\prime}I^{\prime}F^{\prime}m_{F}^{\prime} \rangle \\&= R_{nl,n^{\prime}l^{\prime}}A_{sjIFm_{F},sj^{\prime}IF^{\prime}m_{F}^{\prime}}^{ll^{\prime}}\delta_{l,l^{\prime}\pm1}\delta_{s,s^{\prime}}\delta_{I,I^{\prime}},
\end{split}
\label{Eq:1}
\end{eqnarray}
where $q = m_{F} - m_{F}' $,  \( R_{nl,n^{\prime}l^{\prime}} =\langle nl|\hat{r}|n^{\prime}l^{\prime}\rangle\), the angular contribution $A_{sjIFm_{F},sj^{\prime}IF^{\prime}m_{F}^{\prime}}^{ll^{\prime}}$ reordering by the Wigner $3j$ symbol and Wigner $6j$ symbol reads
\begin{eqnarray}
\begin{split}
A_{sjIFm_{F},sj^{\prime}IF^{\prime}m_{F}^{\prime}}^{ll^{\prime}}=
A_{0}\left( 
\begin{array}{ccc}
F' & 1 & F \\ 
m_{F}'  & q & -m_{F}
\end{array}
\right)\left\{ 
\begin{array}{ccc}
j & j^{\prime } & 1 \\ 
F^{\prime } & F & I
\end{array}
\right\} \\\left\{ 
\begin{array}{ccc}
l & l^{\prime } & 1 \\ 
j^{\prime } & j & s%
\end{array}%
\right\} \text{,}
\label{Eq:2}
\end{split}
\end{eqnarray}  
with $
A_{0}=\sqrt{\left( 2l+1\right)\left( 2j+1\right) \left( 2j^{\prime
}+1\right) \left( 2F+1\right) \left( 2F^{\prime }+1\right) }\\(-1)^{2F'+m_{F}+j+j^{\prime }+I+s+l+1}$. It implies that the transition dipole moment can be simplified by factoring out the dependence on \( I \), \( F \), and \( m_F \) into Wigner symbols and the reduced matrix element \( \mu_{J} \) that depends on the quantum numbers \( n \), \( l \), \( s \), and \( j \) of electronic states. Thus, the transition dipole moment ${\mu}_{FF'}$ can be given by 
\begin{eqnarray}
\begin{split}
\mu_{FF'}=\mu_{J}\left( -1\right) ^{2F^{\prime }+m_{F}+j+I}\sqrt{%
\left( 2F+1\right) \left( 2F^{\prime }+1\right) (2j+1)}\\\left( 
\begin{array}{ccc}
F' & 1 & F \\ 
m_{F}'  & q & -m_{F}
\end{array}%
\right) \left\{ 
\begin{array}{ccc}
j & j^{\prime } & 1 \\ 
F^{\prime } & F & I
\end{array}
\right\},
\label{Eq:3}
\end{split}
\end{eqnarray}
where the reduced matrix element $\mu_J$ takes the value of \(\mu_J = 2.9931 \) a.u. for the transition from \(5^{2}S_{1/2}\) to \( 5^{2}P_{1/2}\) \cite{Rb_date_2024}. As a result, we can obtain the transition dipole moments between hyperfine energy levels $|F\rangle$ and $|F'\rangle$ in \(^{87}\mathrm{Rb}\) atoms by considering the corresponding selection rules.

%


\begin{thebibliography}{71}%
\makeatletter
\providecommand \@ifxundefined [1]{%
 \@ifx{#1\undefined}
}%
\providecommand \@ifnum [1]{%
 \ifnum #1\expandafter \@firstoftwo
 \else \expandafter \@secondoftwo
 \fi
}%
\providecommand \@ifx [1]{%
 \ifx #1\expandafter \@firstoftwo
 \else \expandafter \@secondoftwo
 \fi
}%
\providecommand \natexlab [1]{#1}%
\providecommand \enquote  [1]{``#1''}%
\providecommand \bibnamefont  [1]{#1}%
\providecommand \bibfnamefont [1]{#1}%
\providecommand \citenamefont [1]{#1}%
\providecommand \href@noop [0]{\@secondoftwo}%
\providecommand \href [0]{\begingroup \@sanitize@url \@href}%
\providecommand \@href[1]{\@@startlink{#1}\@@href}%
\providecommand \@@href[1]{\endgroup#1\@@endlink}%
\providecommand \@sanitize@url [0]{\catcode `\\12\catcode `\$12\catcode `\&12\catcode `\#12\catcode `\^12\catcode `\_12\catcode `\%12\relax}%
\providecommand \@@startlink[1]{}%
\providecommand \@@endlink[0]{}%
\providecommand \url  [0]{\begingroup\@sanitize@url \@url }%
\providecommand \@url [1]{\endgroup\@href {#1}{\urlprefix }}%
\providecommand \urlprefix  [0]{URL }%
\providecommand \Eprint [0]{\href }%
\providecommand \doibase [0]{https://doi.org/}%
\providecommand \selectlanguage [0]{\@gobble}%
\providecommand \bibinfo  [0]{\@secondoftwo}%
\providecommand \bibfield  [0]{\@secondoftwo}%
\providecommand \translation [1]{[#1]}%
\providecommand \BibitemOpen [0]{}%
\providecommand \bibitemStop [0]{}%
\providecommand \bibitemNoStop [0]{.\EOS\space}%
\providecommand \EOS [0]{\spacefactor3000\relax}%
\providecommand \BibitemShut  [1]{\csname bibitem#1\endcsname}%
\let\auto@bib@innerbib\@empty
\bibitem [{\citenamefont {Nielsen}\ and\ \citenamefont {Chuang}(2010)}]{Nielsen_book_2010}%
  \BibitemOpen
  \bibfield  {author} {\bibinfo {author} {\bibfnamefont {M.~A.}\ \bibnamefont {Nielsen}}\ and\ \bibinfo {author} {\bibfnamefont {I.~L.}\ \bibnamefont {Chuang}},\ }\href@noop {} {\emph {\bibinfo {title} {Quantum Computation and Quantum Information: 10th Anniversary Edition}}}\ (\bibinfo  {publisher} {Cambridge University Press},\ \bibinfo {address} {Cambridge},\ \bibinfo {year} {2010})\BibitemShut {NoStop}%
\bibitem [{\citenamefont {Wu}\ \emph {et~al.}(2021)\citenamefont {Wu}, \citenamefont {Wang}, \citenamefont {Han}, \citenamefont {Jiang}, \citenamefont {Song}, \citenamefont {Xia}, \citenamefont {Su},\ and\ \citenamefont {Li}}]{WuJinLei_PRApplied_2021}%
  \BibitemOpen
  \bibfield  {author} {\bibinfo {author} {\bibfnamefont {J.-L.}\ \bibnamefont {Wu}}, \bibinfo {author} {\bibfnamefont {Y.}~\bibnamefont {Wang}}, \bibinfo {author} {\bibfnamefont {J.-X.}\ \bibnamefont {Han}}, \bibinfo {author} {\bibfnamefont {Y.}~\bibnamefont {Jiang}}, \bibinfo {author} {\bibfnamefont {J.}~\bibnamefont {Song}}, \bibinfo {author} {\bibfnamefont {Y.}~\bibnamefont {Xia}}, \bibinfo {author} {\bibfnamefont {S.-L.}\ \bibnamefont {Su}},\ and\ \bibinfo {author} {\bibfnamefont {W.}~\bibnamefont {Li}},\ }\bibfield  {title} {\bibinfo {title} {Systematic-error-tolerant multiqubit holonomic entangling gates},\ }\href {https://doi.org/10.1103/PhysRevApplied.16.064031} {\bibfield  {journal} {\bibinfo  {journal} {Phys. Rev. Appl.}\ }\textbf {\bibinfo {volume} {16}},\ \bibinfo {pages} {064031} (\bibinfo {year} {2021})}\BibitemShut {NoStop}%
\bibitem [{\citenamefont {Zhang}\ \emph {et~al.}(2023)\citenamefont {Zhang}, \citenamefont {He}, \citenamefont {Sun}, \citenamefont {Zheng}, \citenamefont {Liu}, \citenamefont {Luo}, \citenamefont {Wang}, \citenamefont {Zhu}, \citenamefont {Qiu}, \citenamefont {Shen}, \citenamefont {Wang}, \citenamefont {Lin}, \citenamefont {Yu}, \citenamefont {Li}, \citenamefont {Xiao}, \citenamefont {Li}, \citenamefont {Yang}, \citenamefont {Jiang}, \citenamefont {Dai}, \citenamefont {Zhou}, \citenamefont {Ma}, \citenamefont {Yuan},\ and\ \citenamefont {Pan}}]{ZhangWeiYong_PRL_2023}%
  \BibitemOpen
  \bibfield  {author} {\bibinfo {author} {\bibfnamefont {W.-Y.}\ \bibnamefont {Zhang}}, \bibinfo {author} {\bibfnamefont {M.-G.}\ \bibnamefont {He}}, \bibinfo {author} {\bibfnamefont {H.}~\bibnamefont {Sun}}, \bibinfo {author} {\bibfnamefont {Y.-G.}\ \bibnamefont {Zheng}}, \bibinfo {author} {\bibfnamefont {Y.}~\bibnamefont {Liu}}, \bibinfo {author} {\bibfnamefont {A.}~\bibnamefont {Luo}}, \bibinfo {author} {\bibfnamefont {H.-Y.}\ \bibnamefont {Wang}}, \bibinfo {author} {\bibfnamefont {Z.-H.}\ \bibnamefont {Zhu}}, \bibinfo {author} {\bibfnamefont {P.-Y.}\ \bibnamefont {Qiu}}, \bibinfo {author} {\bibfnamefont {Y.-C.}\ \bibnamefont {Shen}}, \bibinfo {author} {\bibfnamefont {X.-K.}\ \bibnamefont {Wang}}, \bibinfo {author} {\bibfnamefont {W.}~\bibnamefont {Lin}}, \bibinfo {author} {\bibfnamefont {S.-T.}\ \bibnamefont {Yu}}, \bibinfo {author} {\bibfnamefont {B.-C.}\ \bibnamefont {Li}}, \bibinfo {author} {\bibfnamefont {B.}~\bibnamefont {Xiao}}, \bibinfo {author} {\bibfnamefont {M.-D.}\ \bibnamefont {Li}}, \bibinfo
  {author} {\bibfnamefont {Y.-M.}\ \bibnamefont {Yang}}, \bibinfo {author} {\bibfnamefont {X.}~\bibnamefont {Jiang}}, \bibinfo {author} {\bibfnamefont {H.-N.}\ \bibnamefont {Dai}}, \bibinfo {author} {\bibfnamefont {Y.}~\bibnamefont {Zhou}}, \bibinfo {author} {\bibfnamefont {X.}~\bibnamefont {Ma}}, \bibinfo {author} {\bibfnamefont {Z.-S.}\ \bibnamefont {Yuan}},\ and\ \bibinfo {author} {\bibfnamefont {J.-W.}\ \bibnamefont {Pan}},\ }\bibfield  {title} {\bibinfo {title} {Scalable multipartite entanglement created by spin exchange in an optical lattice},\ }\href {https://doi.org/10.1103/PhysRevLett.131.073401} {\bibfield  {journal} {\bibinfo  {journal} {Phys. Rev. Lett.}\ }\textbf {\bibinfo {volume} {131}},\ \bibinfo {pages} {073401} (\bibinfo {year} {2023})}\BibitemShut {NoStop}%
\bibitem [{\citenamefont {Evered}\ \emph {et~al.}(2023)\citenamefont {Evered}, \citenamefont {Bluvstein}, \citenamefont {Kalinowski}, \citenamefont {Ebadi}, \citenamefont {Manovitz}, \citenamefont {Zhou}, \citenamefont {Li}, \citenamefont {Geim}, \citenamefont {Wang}, \citenamefont {Maskara}, \citenamefont {Levine}, \citenamefont {Semeghini}, \citenamefont {Greiner}, \citenamefont {Vuletić},\ and\ \citenamefont {Lukin}}]{Evered_nature_2023}%
  \BibitemOpen
  \bibfield  {author} {\bibinfo {author} {\bibfnamefont {S.~J.}\ \bibnamefont {Evered}}, \bibinfo {author} {\bibfnamefont {D.}~\bibnamefont {Bluvstein}}, \bibinfo {author} {\bibfnamefont {M.}~\bibnamefont {Kalinowski}}, \bibinfo {author} {\bibfnamefont {S.}~\bibnamefont {Ebadi}}, \bibinfo {author} {\bibfnamefont {T.}~\bibnamefont {Manovitz}}, \bibinfo {author} {\bibfnamefont {H.}~\bibnamefont {Zhou}}, \bibinfo {author} {\bibfnamefont {S.~H.}\ \bibnamefont {Li}}, \bibinfo {author} {\bibfnamefont {A.~A.}\ \bibnamefont {Geim}}, \bibinfo {author} {\bibfnamefont {T.~T.}\ \bibnamefont {Wang}}, \bibinfo {author} {\bibfnamefont {N.}~\bibnamefont {Maskara}}, \bibinfo {author} {\bibfnamefont {H.}~\bibnamefont {Levine}}, \bibinfo {author} {\bibfnamefont {G.}~\bibnamefont {Semeghini}}, \bibinfo {author} {\bibfnamefont {M.}~\bibnamefont {Greiner}}, \bibinfo {author} {\bibfnamefont {V.}~\bibnamefont {Vuletić}},\ and\ \bibinfo {author} {\bibfnamefont {M.~D.}\ \bibnamefont {Lukin}},\ }\bibfield  {title} {\bibinfo {title}
  {High-fidelity parallel entangling gates on a neutral-atom quantum computer},\ }\href {https://doi.org/10.1038/s41586-023-06481-y} {\bibfield  {journal} {\bibinfo  {journal} {Nature}\ }\textbf {\bibinfo {volume} {622}},\ \bibinfo {pages} {268} (\bibinfo {year} {2023})}\BibitemShut {NoStop}%
\bibitem [{\citenamefont {Cao}\ \emph {et~al.}(2023)\citenamefont {Cao}, \citenamefont {Chen}, \citenamefont {Liu}, \citenamefont {Mao}, \citenamefont {Xu}, \citenamefont {Wu},\ and\ \citenamefont {You}}]{CaoJiaHao_PRL_2023}%
  \BibitemOpen
  \bibfield  {author} {\bibinfo {author} {\bibfnamefont {J.-H.}\ \bibnamefont {Cao}}, \bibinfo {author} {\bibfnamefont {F.}~\bibnamefont {Chen}}, \bibinfo {author} {\bibfnamefont {Q.}~\bibnamefont {Liu}}, \bibinfo {author} {\bibfnamefont {T.-W.}\ \bibnamefont {Mao}}, \bibinfo {author} {\bibfnamefont {W.-X.}\ \bibnamefont {Xu}}, \bibinfo {author} {\bibfnamefont {L.-N.}\ \bibnamefont {Wu}},\ and\ \bibinfo {author} {\bibfnamefont {L.}~\bibnamefont {You}},\ }\bibfield  {title} {\bibinfo {title} {Detection of entangled states supported by reinforcement learning},\ }\href {https://doi.org/10.1103/PhysRevLett.131.073201} {\bibfield  {journal} {\bibinfo  {journal} {Phys. Rev. Lett.}\ }\textbf {\bibinfo {volume} {131}},\ \bibinfo {pages} {073201} (\bibinfo {year} {2023})}\BibitemShut {NoStop}%
\bibitem [{\citenamefont {Schäfer}\ \emph {et~al.}(2020)\citenamefont {Schäfer}, \citenamefont {Fukuhara}, \citenamefont {Sugawa}, \citenamefont {Takasu},\ and\ \citenamefont {Takahashi}}]{Schäfer_NRP_2020}%
  \BibitemOpen
  \bibfield  {author} {\bibinfo {author} {\bibfnamefont {F.}~\bibnamefont {Schäfer}}, \bibinfo {author} {\bibfnamefont {T.}~\bibnamefont {Fukuhara}}, \bibinfo {author} {\bibfnamefont {S.}~\bibnamefont {Sugawa}}, \bibinfo {author} {\bibfnamefont {Y.}~\bibnamefont {Takasu}},\ and\ \bibinfo {author} {\bibfnamefont {Y.}~\bibnamefont {Takahashi}},\ }\bibfield  {title} {\bibinfo {title} {Tools for quantum simulation with ultracold atoms in optical lattices},\ }\href {https://doi.org/10.1038/s42254-020-0195-3} {\bibfield  {journal} {\bibinfo  {journal} {Nat. Rev. Phys.}\ }\textbf {\bibinfo {volume} {2}},\ \bibinfo {pages} {411} (\bibinfo {year} {2020})}\BibitemShut {NoStop}%
\bibitem [{\citenamefont {Ye}\ \emph {et~al.}(2023)\citenamefont {Ye}, \citenamefont {Tian}, \citenamefont {Lin}, \citenamefont {Luo}, \citenamefont {You}, \citenamefont {Hu}, \citenamefont {Zhang}, \citenamefont {Chen},\ and\ \citenamefont {Li}}]{Meng_PRL_2023}%
  \BibitemOpen
  \bibfield  {author} {\bibinfo {author} {\bibfnamefont {M.}~\bibnamefont {Ye}}, \bibinfo {author} {\bibfnamefont {Y.}~\bibnamefont {Tian}}, \bibinfo {author} {\bibfnamefont {J.}~\bibnamefont {Lin}}, \bibinfo {author} {\bibfnamefont {Y.}~\bibnamefont {Luo}}, \bibinfo {author} {\bibfnamefont {J.}~\bibnamefont {You}}, \bibinfo {author} {\bibfnamefont {J.}~\bibnamefont {Hu}}, \bibinfo {author} {\bibfnamefont {W.}~\bibnamefont {Zhang}}, \bibinfo {author} {\bibfnamefont {W.}~\bibnamefont {Chen}},\ and\ \bibinfo {author} {\bibfnamefont {X.}~\bibnamefont {Li}},\ }\bibfield  {title} {\bibinfo {title} {Universal quantum optimization with cold atoms in an optical cavity},\ }\href {https://doi.org/10.1103/PhysRevLett.131.103601} {\bibfield  {journal} {\bibinfo  {journal} {Phys. Rev. Lett.}\ }\textbf {\bibinfo {volume} {131}},\ \bibinfo {pages} {103601} (\bibinfo {year} {2023})}\BibitemShut {NoStop}%
\bibitem [{\citenamefont {Chalopin}\ \emph {et~al.}(2025)\citenamefont {Chalopin}, \citenamefont {Bojovi\ifmmode~\acute{c}\else \'{c}\fi{}}, \citenamefont {Bourgund}, \citenamefont {Wang}, \citenamefont {Franz}, \citenamefont {Bloch},\ and\ \citenamefont {Hilker}}]{Chalopin_PRL_2025}%
  \BibitemOpen
  \bibfield  {author} {\bibinfo {author} {\bibfnamefont {T.}~\bibnamefont {Chalopin}}, \bibinfo {author} {\bibfnamefont {P.}~\bibnamefont {Bojovi\ifmmode~\acute{c}\else \'{c}\fi{}}}, \bibinfo {author} {\bibfnamefont {D.}~\bibnamefont {Bourgund}}, \bibinfo {author} {\bibfnamefont {S.}~\bibnamefont {Wang}}, \bibinfo {author} {\bibfnamefont {T.}~\bibnamefont {Franz}}, \bibinfo {author} {\bibfnamefont {I.}~\bibnamefont {Bloch}},\ and\ \bibinfo {author} {\bibfnamefont {T.}~\bibnamefont {Hilker}},\ }\bibfield  {title} {\bibinfo {title} {Optical superlattice for engineering {Hubbard} couplings in quantum simulation},\ }\href {https://doi.org/10.1103/PhysRevLett.134.053402} {\bibfield  {journal} {\bibinfo  {journal} {Phys. Rev. Lett.}\ }\textbf {\bibinfo {volume} {134}},\ \bibinfo {pages} {053402} (\bibinfo {year} {2025})}\BibitemShut {NoStop}%
\bibitem [{\citenamefont {Halimeh}\ \emph {et~al.}(2025)\citenamefont {Halimeh}, \citenamefont {Aidelsburger}, \citenamefont {Grusdt}, \citenamefont {Hauke},\ and\ \citenamefont {Yang}}]{Halimeh_NP_2025}%
  \BibitemOpen
  \bibfield  {author} {\bibinfo {author} {\bibfnamefont {J.~C.}\ \bibnamefont {Halimeh}}, \bibinfo {author} {\bibfnamefont {M.}~\bibnamefont {Aidelsburger}}, \bibinfo {author} {\bibfnamefont {F.}~\bibnamefont {Grusdt}}, \bibinfo {author} {\bibfnamefont {P.}~\bibnamefont {Hauke}},\ and\ \bibinfo {author} {\bibfnamefont {B.}~\bibnamefont {Yang}},\ }\bibfield  {title} {\bibinfo {title} {Cold-atom quantum simulators of gauge theories},\ }\href {https://doi.org/10.1038/s41567-024-02721-8} {\bibfield  {journal} {\bibinfo  {journal} {Nat. Phys.}\ }\textbf {\bibinfo {volume} {21}},\ \bibinfo {pages} {25} (\bibinfo {year} {2025})}\BibitemShut {NoStop}%
\bibitem [{\citenamefont {Zhou}\ \emph {et~al.}(2015)\citenamefont {Zhou}, \citenamefont {Long}, \citenamefont {Tang}, \citenamefont {Chen}, \citenamefont {Gao}, \citenamefont {Peng}, \citenamefont {Duan}, \citenamefont {Zhong}, \citenamefont {Xiong}, \citenamefont {Wang}, \citenamefont {Zhang},\ and\ \citenamefont {Zhan}}]{ZhouLin_PRL_2015}%
  \BibitemOpen
  \bibfield  {author} {\bibinfo {author} {\bibfnamefont {L.}~\bibnamefont {Zhou}}, \bibinfo {author} {\bibfnamefont {S.}~\bibnamefont {Long}}, \bibinfo {author} {\bibfnamefont {B.}~\bibnamefont {Tang}}, \bibinfo {author} {\bibfnamefont {X.}~\bibnamefont {Chen}}, \bibinfo {author} {\bibfnamefont {F.}~\bibnamefont {Gao}}, \bibinfo {author} {\bibfnamefont {W.}~\bibnamefont {Peng}}, \bibinfo {author} {\bibfnamefont {W.}~\bibnamefont {Duan}}, \bibinfo {author} {\bibfnamefont {J.}~\bibnamefont {Zhong}}, \bibinfo {author} {\bibfnamefont {Z.}~\bibnamefont {Xiong}}, \bibinfo {author} {\bibfnamefont {J.}~\bibnamefont {Wang}}, \bibinfo {author} {\bibfnamefont {Y.}~\bibnamefont {Zhang}},\ and\ \bibinfo {author} {\bibfnamefont {M.}~\bibnamefont {Zhan}},\ }\bibfield  {title} {\bibinfo {title} {Test of equivalence principle at $1{0}^{\ensuremath{-}8}$ level by a dual-species double-diffraction \text{Raman} atom interferometer},\ }\href {https://doi.org/10.1103/PhysRevLett.115.013004} {\bibfield  {journal} {\bibinfo
  {journal} {Phys. Rev. Lett.}\ }\textbf {\bibinfo {volume} {115}},\ \bibinfo {pages} {013004} (\bibinfo {year} {2015})}\BibitemShut {NoStop}%
\bibitem [{\citenamefont {Zhou}\ \emph {et~al.}(2021)\citenamefont {Zhou}, \citenamefont {He}, \citenamefont {Yan}, \citenamefont {Chen}, \citenamefont {Gao}, \citenamefont {Duan}, \citenamefont {Ji}, \citenamefont {Xu}, \citenamefont {Tang}, \citenamefont {Zhou}, \citenamefont {Barthwal}, \citenamefont {Wang}, \citenamefont {Hou}, \citenamefont {Xiong}, \citenamefont {Zhang}, \citenamefont {Liu}, \citenamefont {Ni}, \citenamefont {Wang},\ and\ \citenamefont {Zhan}}]{ZhouLin_PRA_2021}%
  \BibitemOpen
  \bibfield  {author} {\bibinfo {author} {\bibfnamefont {L.}~\bibnamefont {Zhou}}, \bibinfo {author} {\bibfnamefont {C.}~\bibnamefont {He}}, \bibinfo {author} {\bibfnamefont {S.-T.}\ \bibnamefont {Yan}}, \bibinfo {author} {\bibfnamefont {X.}~\bibnamefont {Chen}}, \bibinfo {author} {\bibfnamefont {D.-F.}\ \bibnamefont {Gao}}, \bibinfo {author} {\bibfnamefont {W.-T.}\ \bibnamefont {Duan}}, \bibinfo {author} {\bibfnamefont {Y.-H.}\ \bibnamefont {Ji}}, \bibinfo {author} {\bibfnamefont {R.-D.}\ \bibnamefont {Xu}}, \bibinfo {author} {\bibfnamefont {B.}~\bibnamefont {Tang}}, \bibinfo {author} {\bibfnamefont {C.}~\bibnamefont {Zhou}}, \bibinfo {author} {\bibfnamefont {S.}~\bibnamefont {Barthwal}}, \bibinfo {author} {\bibfnamefont {Q.}~\bibnamefont {Wang}}, \bibinfo {author} {\bibfnamefont {Z.}~\bibnamefont {Hou}}, \bibinfo {author} {\bibfnamefont {Z.-Y.}\ \bibnamefont {Xiong}}, \bibinfo {author} {\bibfnamefont {Y.-Z.}\ \bibnamefont {Zhang}}, \bibinfo {author} {\bibfnamefont {M.}~\bibnamefont {Liu}}, \bibinfo {author}
  {\bibfnamefont {W.-T.}\ \bibnamefont {Ni}}, \bibinfo {author} {\bibfnamefont {J.}~\bibnamefont {Wang}},\ and\ \bibinfo {author} {\bibfnamefont {M.-S.}\ \bibnamefont {Zhan}},\ }\bibfield  {title} {\bibinfo {title} {Joint mass-and-energy test of the equivalence principle at the ${10}^{\ensuremath{-}10}$ level using atoms with specified mass and internal energy},\ }\href {https://doi.org/10.1103/PhysRevA.104.022822} {\bibfield  {journal} {\bibinfo  {journal} {Phys. Rev. A}\ }\textbf {\bibinfo {volume} {104}},\ \bibinfo {pages} {022822} (\bibinfo {year} {2021})}\BibitemShut {NoStop}%
\bibitem [{\citenamefont {Anders}\ \emph {et~al.}(2021)\citenamefont {Anders}, \citenamefont {Idel}, \citenamefont {Feldmann}, \citenamefont {Bondarenko}, \citenamefont {Loriani}, \citenamefont {Lange}, \citenamefont {Peise}, \citenamefont {Gersemann}, \citenamefont {Meyer-Hoppe}, \citenamefont {Abend}, \citenamefont {Gaaloul}, \citenamefont {Schubert}, \citenamefont {Schlippert}, \citenamefont {Santos}, \citenamefont {Rasel},\ and\ \citenamefont {Klempt}}]{Anders_PRL_2021}%
  \BibitemOpen
  \bibfield  {author} {\bibinfo {author} {\bibfnamefont {F.}~\bibnamefont {Anders}}, \bibinfo {author} {\bibfnamefont {A.}~\bibnamefont {Idel}}, \bibinfo {author} {\bibfnamefont {P.}~\bibnamefont {Feldmann}}, \bibinfo {author} {\bibfnamefont {D.}~\bibnamefont {Bondarenko}}, \bibinfo {author} {\bibfnamefont {S.}~\bibnamefont {Loriani}}, \bibinfo {author} {\bibfnamefont {K.}~\bibnamefont {Lange}}, \bibinfo {author} {\bibfnamefont {J.}~\bibnamefont {Peise}}, \bibinfo {author} {\bibfnamefont {M.}~\bibnamefont {Gersemann}}, \bibinfo {author} {\bibfnamefont {B.}~\bibnamefont {Meyer-Hoppe}}, \bibinfo {author} {\bibfnamefont {S.}~\bibnamefont {Abend}}, \bibinfo {author} {\bibfnamefont {N.}~\bibnamefont {Gaaloul}}, \bibinfo {author} {\bibfnamefont {C.}~\bibnamefont {Schubert}}, \bibinfo {author} {\bibfnamefont {D.}~\bibnamefont {Schlippert}}, \bibinfo {author} {\bibfnamefont {L.}~\bibnamefont {Santos}}, \bibinfo {author} {\bibfnamefont {E.}~\bibnamefont {Rasel}},\ and\ \bibinfo {author} {\bibfnamefont
  {C.}~\bibnamefont {Klempt}},\ }\bibfield  {title} {\bibinfo {title} {Momentum entanglement for atom interferometry},\ }\href {https://doi.org/10.1103/PhysRevLett.127.140402} {\bibfield  {journal} {\bibinfo  {journal} {Phys. Rev. Lett.}\ }\textbf {\bibinfo {volume} {127}},\ \bibinfo {pages} {140402} (\bibinfo {year} {2021})}\BibitemShut {NoStop}%
\bibitem [{\citenamefont {Yin}\ \emph {et~al.}(2022)\citenamefont {Yin}, \citenamefont {Lu}, \citenamefont {Li}, \citenamefont {Xia}, \citenamefont {Wang}, \citenamefont {Zhang},\ and\ \citenamefont {Chang}}]{YinMoJuan_PRL_2022}%
  \BibitemOpen
  \bibfield  {author} {\bibinfo {author} {\bibfnamefont {M.-J.}\ \bibnamefont {Yin}}, \bibinfo {author} {\bibfnamefont {X.-T.}\ \bibnamefont {Lu}}, \bibinfo {author} {\bibfnamefont {T.}~\bibnamefont {Li}}, \bibinfo {author} {\bibfnamefont {J.-J.}\ \bibnamefont {Xia}}, \bibinfo {author} {\bibfnamefont {T.}~\bibnamefont {Wang}}, \bibinfo {author} {\bibfnamefont {X.-F.}\ \bibnamefont {Zhang}},\ and\ \bibinfo {author} {\bibfnamefont {H.}~\bibnamefont {Chang}},\ }\bibfield  {title} {\bibinfo {title} {Floquet engineering \text{Hz}-level \text{Rabi} spectra in shallow optical lattice clock},\ }\href {https://doi.org/10.1103/PhysRevLett.128.073603} {\bibfield  {journal} {\bibinfo  {journal} {Phys. Rev. Lett.}\ }\textbf {\bibinfo {volume} {128}},\ \bibinfo {pages} {073603} (\bibinfo {year} {2022})}\BibitemShut {NoStop}%
\bibitem [{\citenamefont {Panda}\ \emph {et~al.}(2024)\citenamefont {Panda}, \citenamefont {Tao}, \citenamefont {Egelhoff}, \citenamefont {Ceja}, \citenamefont {Xu},\ and\ \citenamefont {M{\"u}ller}}]{Panda_NP_2024}%
  \BibitemOpen
  \bibfield  {author} {\bibinfo {author} {\bibfnamefont {C.~D.}\ \bibnamefont {Panda}}, \bibinfo {author} {\bibfnamefont {M.}~\bibnamefont {Tao}}, \bibinfo {author} {\bibfnamefont {J.}~\bibnamefont {Egelhoff}}, \bibinfo {author} {\bibfnamefont {M.}~\bibnamefont {Ceja}}, \bibinfo {author} {\bibfnamefont {V.}~\bibnamefont {Xu}},\ and\ \bibinfo {author} {\bibfnamefont {H.}~\bibnamefont {M{\"u}ller}},\ }\bibfield  {title} {\bibinfo {title} {Coherence limits in lattice atom interferometry at the one-minute scale},\ }\href {https://doi.org/10.1038/s41567-024-02518-9} {\bibfield  {journal} {\bibinfo  {journal} {Nat. Phys.}\ }\textbf {\bibinfo {volume} {20}},\ \bibinfo {pages} {1234} (\bibinfo {year} {2024})}\BibitemShut {NoStop}%
\bibitem [{\citenamefont {Böhi}\ \emph {et~al.}(2009)\citenamefont {Böhi}, \citenamefont {Riedel}, \citenamefont {Hoffrogge}, \citenamefont {Reichel}, \citenamefont {Hänsch},\ and\ \citenamefont {Treutlein}}]{Bohi_NP_2009}%
  \BibitemOpen
  \bibfield  {author} {\bibinfo {author} {\bibfnamefont {P.}~\bibnamefont {Böhi}}, \bibinfo {author} {\bibfnamefont {M.~F.}\ \bibnamefont {Riedel}}, \bibinfo {author} {\bibfnamefont {J.}~\bibnamefont {Hoffrogge}}, \bibinfo {author} {\bibfnamefont {J.}~\bibnamefont {Reichel}}, \bibinfo {author} {\bibfnamefont {T.~W.}\ \bibnamefont {Hänsch}},\ and\ \bibinfo {author} {\bibfnamefont {P.}~\bibnamefont {Treutlein}},\ }\bibfield  {title} {\bibinfo {title} {Coherent manipulation of {Bose–Einstein condensates} with state-dependent microwave potentials on an atom chip},\ }\href {https://doi.org/10.1038/nphys1329} {\bibfield  {journal} {\bibinfo  {journal} {Nat. Phys}\ }\textbf {\bibinfo {volume} {5}},\ \bibinfo {pages} {592} (\bibinfo {year} {2009})}\BibitemShut {NoStop}%
\bibitem [{\citenamefont {Riedel}\ \emph {et~al.}(2010)\citenamefont {Riedel}, \citenamefont {Böhi}, \citenamefont {Li}, \citenamefont {Hänsch}, \citenamefont {Sinatra},\ and\ \citenamefont {Treutlein}}]{Riedel_Nature_2010}%
  \BibitemOpen
  \bibfield  {author} {\bibinfo {author} {\bibfnamefont {M.~F.}\ \bibnamefont {Riedel}}, \bibinfo {author} {\bibfnamefont {P.}~\bibnamefont {Böhi}}, \bibinfo {author} {\bibfnamefont {Y.}~\bibnamefont {Li}}, \bibinfo {author} {\bibfnamefont {T.~W.}\ \bibnamefont {Hänsch}}, \bibinfo {author} {\bibfnamefont {A.}~\bibnamefont {Sinatra}},\ and\ \bibinfo {author} {\bibfnamefont {P.}~\bibnamefont {Treutlein}},\ }\bibfield  {title} {\bibinfo {title} {Atom-chip-based generation of entanglement for quantum metrology},\ }\href {https://doi.org/10.1038/nature08988} {\bibfield  {journal} {\bibinfo  {journal} {Nature}\ }\textbf {\bibinfo {volume} {464}},\ \bibinfo {pages} {1170} (\bibinfo {year} {2010})}\BibitemShut {NoStop}%
\bibitem [{\citenamefont {Pershin}\ \emph {et~al.}(2020)\citenamefont {Pershin}, \citenamefont {Yaroshenko}, \citenamefont {Tsyganok}, \citenamefont {Khlebnikov}, \citenamefont {Davletov}, \citenamefont {Shaykin}, \citenamefont {Gadylshin}, \citenamefont {Cojocaru}, \citenamefont {Svechnikov}, \citenamefont {Kapitanova},\ and\ \citenamefont {Akimov}}]{Pershin_(PRA)_2020}%
  \BibitemOpen
  \bibfield  {author} {\bibinfo {author} {\bibfnamefont {D.~A.}\ \bibnamefont {Pershin}}, \bibinfo {author} {\bibfnamefont {V.~V.}\ \bibnamefont {Yaroshenko}}, \bibinfo {author} {\bibfnamefont {V.~V.}\ \bibnamefont {Tsyganok}}, \bibinfo {author} {\bibfnamefont {V.~A.}\ \bibnamefont {Khlebnikov}}, \bibinfo {author} {\bibfnamefont {E.~T.}\ \bibnamefont {Davletov}}, \bibinfo {author} {\bibfnamefont {D.~V.}\ \bibnamefont {Shaykin}}, \bibinfo {author} {\bibfnamefont {E.~R.}\ \bibnamefont {Gadylshin}}, \bibinfo {author} {\bibfnamefont {I.~S.}\ \bibnamefont {Cojocaru}}, \bibinfo {author} {\bibfnamefont {E.~L.}\ \bibnamefont {Svechnikov}}, \bibinfo {author} {\bibfnamefont {P.~V.}\ \bibnamefont {Kapitanova}},\ and\ \bibinfo {author} {\bibfnamefont {A.~V.}\ \bibnamefont {Akimov}},\ }\bibfield  {title} {\bibinfo {title} {Microwave coherent spectroscopy of ultracold thulium atoms},\ }\href {https://doi.org/10.1103/PhysRevA.102.043114} {\bibfield  {journal} {\bibinfo  {journal} {Phys. Rev. A}\ }\textbf {\bibinfo {volume}
  {102}},\ \bibinfo {pages} {043114} (\bibinfo {year} {2020})}\BibitemShut {NoStop}%
\bibitem [{\citenamefont {Kondo}\ \emph {et~al.}(2024)\citenamefont {Kondo}, \citenamefont {Rittenhouse}, \citenamefont {Magalh\~aes}, \citenamefont {Rokaj}, \citenamefont {Mistakidis}, \citenamefont {Sadeghpour},\ and\ \citenamefont {Marcassa}}]{Kondo_PRA_2024}%
  \BibitemOpen
  \bibfield  {author} {\bibinfo {author} {\bibfnamefont {J.~D.~M.}\ \bibnamefont {Kondo}}, \bibinfo {author} {\bibfnamefont {S.~T.}\ \bibnamefont {Rittenhouse}}, \bibinfo {author} {\bibfnamefont {D.~V.}\ \bibnamefont {Magalh\~aes}}, \bibinfo {author} {\bibfnamefont {V.}~\bibnamefont {Rokaj}}, \bibinfo {author} {\bibfnamefont {S.~I.}\ \bibnamefont {Mistakidis}}, \bibinfo {author} {\bibfnamefont {H.~R.}\ \bibnamefont {Sadeghpour}},\ and\ \bibinfo {author} {\bibfnamefont {L.~G.}\ \bibnamefont {Marcassa}},\ }\bibfield  {title} {\bibinfo {title} {Multiphoton-dressed \text{Rydberg} excitations in a microwave cavity with ultracold \text{Rb} atoms},\ }\href {https://doi.org/10.1103/PhysRevA.110.L061301} {\bibfield  {journal} {\bibinfo  {journal} {Phys. Rev. A}\ }\textbf {\bibinfo {volume} {110}},\ \bibinfo {pages} {L061301} (\bibinfo {year} {2024})}\BibitemShut {NoStop}%
\bibitem [{\citenamefont {Condon}\ \emph {et~al.}(2019)\citenamefont {Condon}, \citenamefont {Rabault}, \citenamefont {Barrett}, \citenamefont {Chichet}, \citenamefont {Arguel}, \citenamefont {Eneriz-Imaz}, \citenamefont {Naik}, \citenamefont {Bertoldi}, \citenamefont {Battelier}, \citenamefont {Bouyer},\ and\ \citenamefont {Landragin}}]{Condon_PRL_(2019)}%
  \BibitemOpen
  \bibfield  {author} {\bibinfo {author} {\bibfnamefont {G.}~\bibnamefont {Condon}}, \bibinfo {author} {\bibfnamefont {M.}~\bibnamefont {Rabault}}, \bibinfo {author} {\bibfnamefont {B.}~\bibnamefont {Barrett}}, \bibinfo {author} {\bibfnamefont {L.}~\bibnamefont {Chichet}}, \bibinfo {author} {\bibfnamefont {R.}~\bibnamefont {Arguel}}, \bibinfo {author} {\bibfnamefont {H.}~\bibnamefont {Eneriz-Imaz}}, \bibinfo {author} {\bibfnamefont {D.}~\bibnamefont {Naik}}, \bibinfo {author} {\bibfnamefont {A.}~\bibnamefont {Bertoldi}}, \bibinfo {author} {\bibfnamefont {B.}~\bibnamefont {Battelier}}, \bibinfo {author} {\bibfnamefont {P.}~\bibnamefont {Bouyer}},\ and\ \bibinfo {author} {\bibfnamefont {A.}~\bibnamefont {Landragin}},\ }\bibfield  {title} {\bibinfo {title} {All-optical \text{Bose-Einstein} condensates in microgravity},\ }\href {https://doi.org/10.1103/PhysRevLett.123.240402} {\bibfield  {journal} {\bibinfo  {journal} {Phys. Rev. Lett.}\ }\textbf {\bibinfo {volume} {123}},\ \bibinfo {pages} {240402} (\bibinfo
  {year} {2019})}\BibitemShut {NoStop}%
\bibitem [{\citenamefont {Arunkumar}\ \emph {et~al.}(2019)\citenamefont {Arunkumar}, \citenamefont {Jagannathan},\ and\ \citenamefont {Thomas}}]{Arunkumar_PRL_2019}%
  \BibitemOpen
  \bibfield  {author} {\bibinfo {author} {\bibfnamefont {N.}~\bibnamefont {Arunkumar}}, \bibinfo {author} {\bibfnamefont {A.}~\bibnamefont {Jagannathan}},\ and\ \bibinfo {author} {\bibfnamefont {J.~E.}\ \bibnamefont {Thomas}},\ }\bibfield  {title} {\bibinfo {title} {Designer spatial control of interactions in ultracold gases},\ }\href {https://doi.org/10.1103/PhysRevLett.122.040405} {\bibfield  {journal} {\bibinfo  {journal} {Phys. Rev. Lett.}\ }\textbf {\bibinfo {volume} {122}},\ \bibinfo {pages} {040405} (\bibinfo {year} {2019})}\BibitemShut {NoStop}%
\bibitem [{\citenamefont {Guo}\ \emph {et~al.}(2019{\natexlab{a}})\citenamefont {Guo}, \citenamefont {Luo}, \citenamefont {Ma},\ and\ \citenamefont {Shu}}]{Chuan-Cun_PRA_2019}%
  \BibitemOpen
  \bibfield  {author} {\bibinfo {author} {\bibfnamefont {Y.}~\bibnamefont {Guo}}, \bibinfo {author} {\bibfnamefont {X.}~\bibnamefont {Luo}}, \bibinfo {author} {\bibfnamefont {S.}~\bibnamefont {Ma}},\ and\ \bibinfo {author} {\bibfnamefont {C.-C.}\ \bibnamefont {Shu}},\ }\bibfield  {title} {\bibinfo {title} {All-optical generation of quantum entangled states with strictly constrained ultrafast laser pulses},\ }\href {https://doi.org/10.1103/PhysRevA.100.023409} {\bibfield  {journal} {\bibinfo  {journal} {Phys. Rev. A}\ }\textbf {\bibinfo {volume} {100}},\ \bibinfo {pages} {023409} (\bibinfo {year} {2019}{\natexlab{a}})}\BibitemShut {NoStop}%
\bibitem [{\citenamefont {Weckesser}\ \emph {et~al.}(2021)\citenamefont {Weckesser}, \citenamefont {Thielemann}, \citenamefont {Wiater}, \citenamefont {Wojciechowska}, \citenamefont {Karpa}, \citenamefont {Jachymski}, \citenamefont {Tomza}, \citenamefont {Walker},\ and\ \citenamefont {Schaetz}}]{Weckesser_Nature_2021}%
  \BibitemOpen
  \bibfield  {author} {\bibinfo {author} {\bibfnamefont {P.}~\bibnamefont {Weckesser}}, \bibinfo {author} {\bibfnamefont {F.}~\bibnamefont {Thielemann}}, \bibinfo {author} {\bibfnamefont {D.}~\bibnamefont {Wiater}}, \bibinfo {author} {\bibfnamefont {A.}~\bibnamefont {Wojciechowska}}, \bibinfo {author} {\bibfnamefont {L.}~\bibnamefont {Karpa}}, \bibinfo {author} {\bibfnamefont {K.}~\bibnamefont {Jachymski}}, \bibinfo {author} {\bibfnamefont {M.}~\bibnamefont {Tomza}}, \bibinfo {author} {\bibfnamefont {T.}~\bibnamefont {Walker}},\ and\ \bibinfo {author} {\bibfnamefont {T.}~\bibnamefont {Schaetz}},\ }\bibfield  {title} {\bibinfo {title} {Observation of {Feshbach} resonances between a single ion and ultracold atoms},\ }\href {https://doi.org/10.1038/s41586-021-04112-y} {\bibfield  {journal} {\bibinfo  {journal} {Nature}\ }\textbf {\bibinfo {volume} {600}},\ \bibinfo {pages} {429} (\bibinfo {year} {2021})}\BibitemShut {NoStop}%
\bibitem [{\citenamefont {Herbst}\ \emph {et~al.}(2022)\citenamefont {Herbst}, \citenamefont {Albers}, \citenamefont {Stolzenberg}, \citenamefont {Bode},\ and\ \citenamefont {Schlippert}}]{Herbst_PRA_2022}%
  \BibitemOpen
  \bibfield  {author} {\bibinfo {author} {\bibfnamefont {A.}~\bibnamefont {Herbst}}, \bibinfo {author} {\bibfnamefont {H.}~\bibnamefont {Albers}}, \bibinfo {author} {\bibfnamefont {K.}~\bibnamefont {Stolzenberg}}, \bibinfo {author} {\bibfnamefont {S.}~\bibnamefont {Bode}},\ and\ \bibinfo {author} {\bibfnamefont {D.}~\bibnamefont {Schlippert}},\ }\bibfield  {title} {\bibinfo {title} {Rapid generation of all-optical $^{39}\mathrm{K}$ \text{Bose-Einstein} condensates using a low-field \text{Feshbach} resonance},\ }\href {https://doi.org/10.1103/PhysRevA.106.043320} {\bibfield  {journal} {\bibinfo  {journal} {Phys. Rev. A}\ }\textbf {\bibinfo {volume} {106}},\ \bibinfo {pages} {043320} (\bibinfo {year} {2022})}\BibitemShut {NoStop}%
\bibitem [{\citenamefont {Guo}\ \emph {et~al.}(2019{\natexlab{b}})\citenamefont {Guo}, \citenamefont {Shu}, \citenamefont {Dong},\ and\ \citenamefont {Nori}}]{RRS_PRL_(2019)}%
  \BibitemOpen
  \bibfield  {author} {\bibinfo {author} {\bibfnamefont {Y.}~\bibnamefont {Guo}}, \bibinfo {author} {\bibfnamefont {C.-C.}\ \bibnamefont {Shu}}, \bibinfo {author} {\bibfnamefont {D.}~\bibnamefont {Dong}},\ and\ \bibinfo {author} {\bibfnamefont {F.}~\bibnamefont {Nori}},\ }\bibfield  {title} {\bibinfo {title} {Vanishing and revival of resonance \text{Raman} scattering},\ }\href {https://doi.org/10.1103/PhysRevLett.123.223202} {\bibfield  {journal} {\bibinfo  {journal} {Phys. Rev. Lett.}\ }\textbf {\bibinfo {volume} {123}},\ \bibinfo {pages} {223202} (\bibinfo {year} {2019}{\natexlab{b}})}\BibitemShut {NoStop}%
\bibitem [{\citenamefont {Pucher}\ \emph {et~al.}(2024)\citenamefont {Pucher}, \citenamefont {Kl\"usener}, \citenamefont {Spriestersbach}, \citenamefont {Geiger}, \citenamefont {Schindewolf}, \citenamefont {Bloch},\ and\ \citenamefont {Blatt}}]{Pucher_PRL_2024}%
  \BibitemOpen
  \bibfield  {author} {\bibinfo {author} {\bibfnamefont {S.}~\bibnamefont {Pucher}}, \bibinfo {author} {\bibfnamefont {V.}~\bibnamefont {Kl\"usener}}, \bibinfo {author} {\bibfnamefont {F.}~\bibnamefont {Spriestersbach}}, \bibinfo {author} {\bibfnamefont {J.}~\bibnamefont {Geiger}}, \bibinfo {author} {\bibfnamefont {A.}~\bibnamefont {Schindewolf}}, \bibinfo {author} {\bibfnamefont {I.}~\bibnamefont {Bloch}},\ and\ \bibinfo {author} {\bibfnamefont {S.}~\bibnamefont {Blatt}},\ }\bibfield  {title} {\bibinfo {title} {Fine-structure qubit encoded in metastable strontium trapped in an optical lattice},\ }\href {https://doi.org/10.1103/PhysRevLett.132.150605} {\bibfield  {journal} {\bibinfo  {journal} {Phys. Rev. Lett.}\ }\textbf {\bibinfo {volume} {132}},\ \bibinfo {pages} {150605} (\bibinfo {year} {2024})}\BibitemShut {NoStop}%
\bibitem [{\citenamefont {Unnikrishnan}\ \emph {et~al.}(2024)\citenamefont {Unnikrishnan}, \citenamefont {Ilzh\"ofer}, \citenamefont {Scholz}, \citenamefont {H\"olzl}, \citenamefont {G\"otzelmann}, \citenamefont {Gupta}, \citenamefont {Zhao}, \citenamefont {Krauter}, \citenamefont {Weber}, \citenamefont {Makki}, \citenamefont {B\"uchler}, \citenamefont {Pfau},\ and\ \citenamefont {Meinert}}]{Unnikrishnan_PRL_2024}%
  \BibitemOpen
  \bibfield  {author} {\bibinfo {author} {\bibfnamefont {G.}~\bibnamefont {Unnikrishnan}}, \bibinfo {author} {\bibfnamefont {P.}~\bibnamefont {Ilzh\"ofer}}, \bibinfo {author} {\bibfnamefont {A.}~\bibnamefont {Scholz}}, \bibinfo {author} {\bibfnamefont {C.}~\bibnamefont {H\"olzl}}, \bibinfo {author} {\bibfnamefont {A.}~\bibnamefont {G\"otzelmann}}, \bibinfo {author} {\bibfnamefont {R.~K.}\ \bibnamefont {Gupta}}, \bibinfo {author} {\bibfnamefont {J.}~\bibnamefont {Zhao}}, \bibinfo {author} {\bibfnamefont {J.}~\bibnamefont {Krauter}}, \bibinfo {author} {\bibfnamefont {S.}~\bibnamefont {Weber}}, \bibinfo {author} {\bibfnamefont {N.}~\bibnamefont {Makki}}, \bibinfo {author} {\bibfnamefont {H.~P.}\ \bibnamefont {B\"uchler}}, \bibinfo {author} {\bibfnamefont {T.}~\bibnamefont {Pfau}},\ and\ \bibinfo {author} {\bibfnamefont {F.}~\bibnamefont {Meinert}},\ }\bibfield  {title} {\bibinfo {title} {Coherent control of the fine-structure qubit in a single alkaline-earth atom},\ }\href
  {https://doi.org/10.1103/PhysRevLett.132.150606} {\bibfield  {journal} {\bibinfo  {journal} {Phys. Rev. Lett.}\ }\textbf {\bibinfo {volume} {132}},\ \bibinfo {pages} {150606} (\bibinfo {year} {2024})}\BibitemShut {NoStop}%
\bibitem [{\citenamefont {Lvovsky}\ \emph {et~al.}(2009)\citenamefont {Lvovsky}, \citenamefont {Sanders},\ and\ \citenamefont {Tittel}}]{Lvovsky_NaturePhotonics_2009}%
  \BibitemOpen
  \bibfield  {author} {\bibinfo {author} {\bibfnamefont {A.~I.}\ \bibnamefont {Lvovsky}}, \bibinfo {author} {\bibfnamefont {B.~C.}\ \bibnamefont {Sanders}},\ and\ \bibinfo {author} {\bibfnamefont {W.}~\bibnamefont {Tittel}},\ }\bibfield  {title} {\bibinfo {title} {Optical quantum memory},\ }\href {https://doi.org/10.1038/nphoton.2009.231} {\bibfield  {journal} {\bibinfo  {journal} {Nat. Photonics}\ }\textbf {\bibinfo {volume} {3}},\ \bibinfo {pages} {706} (\bibinfo {year} {2009})}\BibitemShut {NoStop}%
\bibitem [{\citenamefont {Heshami}\ \emph {et~al.}(2016)\citenamefont {Heshami}, \citenamefont {England}, \citenamefont {Humphreys}, \citenamefont {Bustard}, \citenamefont {Acosta}, \citenamefont {Nunn},\ and\ \citenamefont {Sussman}}]{Heshami_JOMP_2016}%
  \BibitemOpen
  \bibfield  {author} {\bibinfo {author} {\bibfnamefont {K.}~\bibnamefont {Heshami}}, \bibinfo {author} {\bibfnamefont {D.}~\bibnamefont {England}}, \bibinfo {author} {\bibfnamefont {P.}~\bibnamefont {Humphreys}}, \bibinfo {author} {\bibfnamefont {P.}~\bibnamefont {Bustard}}, \bibinfo {author} {\bibfnamefont {V.}~\bibnamefont {Acosta}}, \bibinfo {author} {\bibfnamefont {J.}~\bibnamefont {Nunn}},\ and\ \bibinfo {author} {\bibfnamefont {B.}~\bibnamefont {Sussman}},\ }\bibfield  {title} {\bibinfo {title} {Quantum memories: emerging applications and recent advances},\ }\href {https://doi.org/10.1080/09500340.2016.1148212} {\bibfield  {journal} {\bibinfo  {journal} {J. Mod. Opt.}\ }\textbf {\bibinfo {volume} {63}},\ \bibinfo {pages} {2005} (\bibinfo {year} {2016})}\BibitemShut {NoStop}%
\bibitem [{\citenamefont {Wolfowicz}\ and\ \citenamefont {Morton}(2016)}]{Wolfowicz_book_2016}%
  \BibitemOpen
\bibfield  {author} {\bibinfo {author} {\bibfnamefont {G.}~\bibnamefont {Wolfowicz}}%
\ and\ \bibinfo {author} {\bibfnamefont {J.~J.~L.}\ \bibnamefont {Morton}}, }%
\bibinfo {title} {Pulse techniques for quantum information processing},\ in\ \href{https://doi.org/10.1002/9780470034590.emrstm1521}%
{\emph {\bibinfo {booktitle} {eMagRes}}}\
(\bibinfo  {publisher} {John Wiley \& Sons, Ltd},\ \bibinfo {year} {2016})%
\BibitemShut {NoStop}%
\bibitem [{\citenamefont {Gaubatz}\ \emph {et~al.}(1990)\citenamefont {Gaubatz}, \citenamefont {Rudecki}, \citenamefont {Schiemann},\ and\ \citenamefont {Bergmann}}]{Gaubatz_JCP_1990}%
  \BibitemOpen
  \bibfield  {author} {\bibinfo {author} {\bibfnamefont {U.}~\bibnamefont {Gaubatz}}, \bibinfo {author} {\bibfnamefont {P.}~\bibnamefont {Rudecki}}, \bibinfo {author} {\bibfnamefont {S.}~\bibnamefont {Schiemann}},\ and\ \bibinfo {author} {\bibfnamefont {K.}~\bibnamefont {Bergmann}},\ }\bibfield  {title} {\bibinfo {title} {Population transfer between molecular vibrational levels by stimulated \text{Raman} scattering with partially overlapping laser fields. \text{A} new concept and experimental results},\ }\href {https://doi.org/10.1063/1.458514} {\bibfield  {journal} {\bibinfo  {journal} {J. Chem. Phys.}\ }\textbf {\bibinfo {volume} {92}},\ \bibinfo {pages} {5363} (\bibinfo {year} {1990})}\BibitemShut {NoStop}%
\bibitem [{\citenamefont {Shu}\ \emph {et~al.}(2009)\citenamefont {Shu}, \citenamefont {Yu}, \citenamefont {Yuan}, \citenamefont {Hu}, \citenamefont {Yang},\ and\ \citenamefont {Cong}}]{ShuChuan-Cun_PRA_2009}%
  \BibitemOpen
  \bibfield  {author} {\bibinfo {author} {\bibfnamefont {C.-C.}\ \bibnamefont {Shu}}, \bibinfo {author} {\bibfnamefont {J.}~\bibnamefont {Yu}}, \bibinfo {author} {\bibfnamefont {K.-J.}\ \bibnamefont {Yuan}}, \bibinfo {author} {\bibfnamefont {W.-H.}\ \bibnamefont {Hu}}, \bibinfo {author} {\bibfnamefont {J.}~\bibnamefont {Yang}},\ and\ \bibinfo {author} {\bibfnamefont {S.-L.}\ \bibnamefont {Cong}},\ }\bibfield  {title} {\bibinfo {title} {Stimulated \text{Raman} adiabatic passage in molecular electronic states},\ }\href {https://doi.org/10.1103/PhysRevA.79.023418} {\bibfield  {journal} {\bibinfo  {journal} {Phys. Rev. A}\ }\textbf {\bibinfo {volume} {79}},\ \bibinfo {pages} {023418} (\bibinfo {year} {2009})}\BibitemShut {NoStop}%
\bibitem [{\citenamefont {Bergmann}\ \emph {et~al.}(2015)\citenamefont {Bergmann}, \citenamefont {Vitanov},\ and\ \citenamefont {Shore}}]{Bergmann_JCP_2015}%
  \BibitemOpen
  \bibfield  {author} {\bibinfo {author} {\bibfnamefont {K.}~\bibnamefont {Bergmann}}, \bibinfo {author} {\bibfnamefont {N.~V.}\ \bibnamefont {Vitanov}},\ and\ \bibinfo {author} {\bibfnamefont {B.~W.}\ \bibnamefont {Shore}},\ }\bibfield  {title} {\bibinfo {title} {Perspective: Stimulated \text{Raman} adiabatic passage: The status after 25 years},\ }\href {https://doi.org/10.1063/1.4916903} {\bibfield  {journal} {\bibinfo  {journal} {J. Chem. Phys.}\ }\textbf {\bibinfo {volume} {142}},\ \bibinfo {pages} {170901} (\bibinfo {year} {2015})}\BibitemShut {NoStop}%
\bibitem [{\citenamefont {Vitanov}\ \emph {et~al.}(2017)\citenamefont {Vitanov}, \citenamefont {Rangelov}, \citenamefont {Shore},\ and\ \citenamefont {Bergmann}}]{Vitanov_RMP_2017}%
  \BibitemOpen
  \bibfield  {author} {\bibinfo {author} {\bibfnamefont {N.~V.}\ \bibnamefont {Vitanov}}, \bibinfo {author} {\bibfnamefont {A.~A.}\ \bibnamefont {Rangelov}}, \bibinfo {author} {\bibfnamefont {B.~W.}\ \bibnamefont {Shore}},\ and\ \bibinfo {author} {\bibfnamefont {K.}~\bibnamefont {Bergmann}},\ }\bibfield  {title} {\bibinfo {title} {Stimulated \text{Raman} adiabatic passage in physics, chemistry, and beyond},\ }\href {https://doi.org/10.1103/RevModPhys.89.015006} {\bibfield  {journal} {\bibinfo  {journal} {Rev. Mod. Phys.}\ }\textbf {\bibinfo {volume} {89}},\ \bibinfo {pages} {015006} (\bibinfo {year} {2017})}\BibitemShut {NoStop}%
\bibitem [{\citenamefont {Maddox}\ \emph {et~al.}(2024)\citenamefont {Maddox}, \citenamefont {Mortlock}, \citenamefont {Hepworth}, \citenamefont {Raghuram}, \citenamefont {Gregory}, \citenamefont {Guttridge},\ and\ \citenamefont {Cornish}}]{Maddox_PRL_2024}%
  \BibitemOpen
  \bibfield  {author} {\bibinfo {author} {\bibfnamefont {B.~P.}\ \bibnamefont {Maddox}}, \bibinfo {author} {\bibfnamefont {J.~M.}\ \bibnamefont {Mortlock}}, \bibinfo {author} {\bibfnamefont {T.~R.}\ \bibnamefont {Hepworth}}, \bibinfo {author} {\bibfnamefont {A.~P.}\ \bibnamefont {Raghuram}}, \bibinfo {author} {\bibfnamefont {P.~D.}\ \bibnamefont {Gregory}}, \bibinfo {author} {\bibfnamefont {A.}~\bibnamefont {Guttridge}},\ and\ \bibinfo {author} {\bibfnamefont {S.~L.}\ \bibnamefont {Cornish}},\ }\bibfield  {title} {\bibinfo {title} {Enhanced quantum state transfer via feedforward cancellation of optical phase noise},\ }\href {https://doi.org/10.1103/PhysRevLett.133.253202} {\bibfield  {journal} {\bibinfo  {journal} {Phys. Rev. Lett.}\ }\textbf {\bibinfo {volume} {133}},\ \bibinfo {pages} {253202} (\bibinfo {year} {2024})}\BibitemShut {NoStop}%
\bibitem [{\citenamefont {Messiah}(1965)}]{Messiah1965QuantumMV}%
  \BibitemOpen
  \bibfield  {author} {\bibinfo {author} {\bibfnamefont {A.}~\bibnamefont {Messiah}},\ }\href@noop {} {\emph {\bibinfo {title} {Quantum Mechanics}}},\ Vol.~\bibinfo {volume} {II}\ (\bibinfo  {publisher} {North-Holland Publishing Co.},\ \bibinfo {address} {Amsterdam},\ \bibinfo {year} {1965})\BibitemShut {NoStop}%
\bibitem [{\citenamefont {Peterson}\ and\ \citenamefont {Cantrell}(1985)}]{Peterson_PRA_(1985)}%
  \BibitemOpen
  \bibfield  {author} {\bibinfo {author} {\bibfnamefont {G.~L.}\ \bibnamefont {Peterson}}\ and\ \bibinfo {author} {\bibfnamefont {C.~D.}\ \bibnamefont {Cantrell}},\ }\bibfield  {title} {\bibinfo {title} {Adiabatic excitation of multilevel systems},\ }\href {https://doi.org/10.1103/PhysRevA.31.807} {\bibfield  {journal} {\bibinfo  {journal} {Phys. Rev. A}\ }\textbf {\bibinfo {volume} {31}},\ \bibinfo {pages} {807} (\bibinfo {year} {1985})}\BibitemShut {NoStop}%
\bibitem [{\citenamefont {Kuklinski}\ \emph {et~al.}(1989)\citenamefont {Kuklinski}, \citenamefont {Gaubatz}, \citenamefont {Hioe},\ and\ \citenamefont {Bergmann}}]{Kuklinski_PRA_(1989)}%
  \BibitemOpen
  \bibfield  {author} {\bibinfo {author} {\bibfnamefont {J.~R.}\ \bibnamefont {Kuklinski}}, \bibinfo {author} {\bibfnamefont {U.}~\bibnamefont {Gaubatz}}, \bibinfo {author} {\bibfnamefont {F.~T.}\ \bibnamefont {Hioe}},\ and\ \bibinfo {author} {\bibfnamefont {K.}~\bibnamefont {Bergmann}},\ }\bibfield  {title} {\bibinfo {title} {Adiabatic population transfer in a three-level system driven by delayed laser pulses},\ }\href {https://doi.org/10.1103/PhysRevA.40.6741} {\bibfield  {journal} {\bibinfo  {journal} {Phys. Rev. A}\ }\textbf {\bibinfo {volume} {40}},\ \bibinfo {pages} {6741} (\bibinfo {year} {1989})}\BibitemShut {NoStop}%
\bibitem [{\citenamefont {Wang}\ and\ \citenamefont {Plenio}(2016)}]{NASC_PRA_(2016)}%
  \BibitemOpen
  \bibfield  {author} {\bibinfo {author} {\bibfnamefont {Z.-Y.}\ \bibnamefont {Wang}}\ and\ \bibinfo {author} {\bibfnamefont {M.~B.}\ \bibnamefont {Plenio}},\ }\bibfield  {title} {\bibinfo {title} {Necessary and sufficient condition for quantum adiabatic evolution by unitary control fields},\ }\href {https://doi.org/10.1103/PhysRevA.93.052107} {\bibfield  {journal} {\bibinfo  {journal} {Phys. Rev. A}\ }\textbf {\bibinfo {volume} {93}},\ \bibinfo {pages} {052107} (\bibinfo {year} {2016})}\BibitemShut {NoStop}%
\bibitem [{\citenamefont {Marzlin}\ and\ \citenamefont {Sanders}(2004)}]{Marzlin_PRL_2004}%
  \BibitemOpen
  \bibfield  {author} {\bibinfo {author} {\bibfnamefont {K.-P.}\ \bibnamefont {Marzlin}}\ and\ \bibinfo {author} {\bibfnamefont {B.~C.}\ \bibnamefont {Sanders}},\ }\bibfield  {title} {\bibinfo {title} {Inconsistency in the application of the adiabatic theorem},\ }\href {https://doi.org/10.1103/PhysRevLett.93.160408} {\bibfield  {journal} {\bibinfo  {journal} {Phys. Rev. Lett.}\ }\textbf {\bibinfo {volume} {93}},\ \bibinfo {pages} {160408} (\bibinfo {year} {2004})}\BibitemShut {NoStop}%
\bibitem [{\citenamefont {Tong}\ \emph {et~al.}(2007)\citenamefont {Tong}, \citenamefont {Singh}, \citenamefont {Kwek},\ and\ \citenamefont {Oh}}]{Tong_PRL_2007}%
  \BibitemOpen
  \bibfield  {author} {\bibinfo {author} {\bibfnamefont {D.~M.}\ \bibnamefont {Tong}}, \bibinfo {author} {\bibfnamefont {K.}~\bibnamefont {Singh}}, \bibinfo {author} {\bibfnamefont {L.~C.}\ \bibnamefont {Kwek}},\ and\ \bibinfo {author} {\bibfnamefont {C.~H.}\ \bibnamefont {Oh}},\ }\bibfield  {title} {\bibinfo {title} {Sufficiency criterion for the validity of the adiabatic approximation},\ }\href {https://doi.org/10.1103/PhysRevLett.98.150402} {\bibfield  {journal} {\bibinfo  {journal} {Phys. Rev. Lett.}\ }\textbf {\bibinfo {volume} {98}},\ \bibinfo {pages} {150402} (\bibinfo {year} {2007})}\BibitemShut {NoStop}%
\bibitem [{\citenamefont {Boixo}\ and\ \citenamefont {Somma}(2010)}]{Boixo_PRA_2010}%
  \BibitemOpen
  \bibfield  {author} {\bibinfo {author} {\bibfnamefont {S.}~\bibnamefont {Boixo}}\ and\ \bibinfo {author} {\bibfnamefont {R.~D.}\ \bibnamefont {Somma}},\ }\bibfield  {title} {\bibinfo {title} {Necessary condition for the quantum adiabatic approximation},\ }\href {https://doi.org/10.1103/PhysRevA.81.032308} {\bibfield  {journal} {\bibinfo  {journal} {Phys. Rev. A}\ }\textbf {\bibinfo {volume} {81}},\ \bibinfo {pages} {032308} (\bibinfo {year} {2010})}\BibitemShut {NoStop}%
\bibitem [{\citenamefont {Zheng}\ \emph {et~al.}(2022)\citenamefont {Zheng}, \citenamefont {Xu}, \citenamefont {Wang}, \citenamefont {Dong}, \citenamefont {Lan}, \citenamefont {Tan},\ and\ \citenamefont {Yu}}]{ZhengWen_PRAppl_(2022)}%
  \BibitemOpen
  \bibfield  {author} {\bibinfo {author} {\bibfnamefont {W.}~\bibnamefont {Zheng}}, \bibinfo {author} {\bibfnamefont {J.}~\bibnamefont {Xu}}, \bibinfo {author} {\bibfnamefont {Z.}~\bibnamefont {Wang}}, \bibinfo {author} {\bibfnamefont {Y.}~\bibnamefont {Dong}}, \bibinfo {author} {\bibfnamefont {D.}~\bibnamefont {Lan}}, \bibinfo {author} {\bibfnamefont {X.}~\bibnamefont {Tan}},\ and\ \bibinfo {author} {\bibfnamefont {Y.}~\bibnamefont {Yu}},\ }\bibfield  {title} {\bibinfo {title} {Accelerated quantum adiabatic transfer in superconducting qubits},\ }\href {https://doi.org/10.1103/PhysRevApplied.18.044014} {\bibfield  {journal} {\bibinfo  {journal} {Phys. Rev. Appl.}\ }\textbf {\bibinfo {volume} {18}},\ \bibinfo {pages} {044014} (\bibinfo {year} {2022})}\BibitemShut {NoStop}%
\bibitem [{\citenamefont {Liu}\ and\ \citenamefont {Wang}()}]{WangZhen-Yu_arXiv_(2023)}%
  \BibitemOpen
  \bibfield  {author} {\bibinfo {author} {\bibfnamefont {Y.}~\bibnamefont {Liu}}\ and\ \bibinfo {author} {\bibfnamefont {Z.-Y.}\ \bibnamefont {Wang}},\ }\bibfield  {title} {\bibinfo {title} {Shortcuts to adiabaticity with inherent robustness and without auxiliary control},\ }\href {https://arxiv.org/abs/2211.02543} {\ }\Eprint {https://arxiv.org/abs/2211.02543} {arXiv:2211.02543} \BibitemShut {NoStop}%
\bibitem [{\citenamefont {Gong}\ \emph {et~al.}(2023)\citenamefont {Gong}, \citenamefont {Yu}, \citenamefont {Betzholz}, \citenamefont {Chu}, \citenamefont {Yang}, \citenamefont {Wang},\ and\ \citenamefont {Cai}}]{GongMusang_PRA_(2023)}%
  \BibitemOpen
  \bibfield  {author} {\bibinfo {author} {\bibfnamefont {M.}~\bibnamefont {Gong}}, \bibinfo {author} {\bibfnamefont {M.}~\bibnamefont {Yu}}, \bibinfo {author} {\bibfnamefont {R.}~\bibnamefont {Betzholz}}, \bibinfo {author} {\bibfnamefont {Y.}~\bibnamefont {Chu}}, \bibinfo {author} {\bibfnamefont {P.}~\bibnamefont {Yang}}, \bibinfo {author} {\bibfnamefont {Z.}~\bibnamefont {Wang}},\ and\ \bibinfo {author} {\bibfnamefont {J.}~\bibnamefont {Cai}},\ }\bibfield  {title} {\bibinfo {title} {Accelerated quantum control in a three-level system by jumping along the geodesics},\ }\href {https://doi.org/10.1103/PhysRevA.107.L040602} {\bibfield  {journal} {\bibinfo  {journal} {Phys. Rev. A}\ }\textbf {\bibinfo {volume} {107}},\ \bibinfo {pages} {L040602} (\bibinfo {year} {2023})}\BibitemShut {NoStop}%
\bibitem [{\citenamefont {Chen}\ \emph {et~al.}(2024)\citenamefont {Chen}, \citenamefont {Lu}, \citenamefont {Chen},\ and\ \citenamefont {Wang}}]{WangZhen-Yu_PRA_2024}%
  \BibitemOpen
  \bibfield  {author} {\bibinfo {author} {\bibfnamefont {C.}~\bibnamefont {Chen}}, \bibinfo {author} {\bibfnamefont {J.-Y.}\ \bibnamefont {Lu}}, \bibinfo {author} {\bibfnamefont {X.-Y.}\ \bibnamefont {Chen}},\ and\ \bibinfo {author} {\bibfnamefont {Z.-Y.}\ \bibnamefont {Wang}},\ }\bibfield  {title} {\bibinfo {title} {Fast adiabatic preparation of multisqueezed states by jumping along the path},\ }\href {https://doi.org/10.1103/PhysRevA.110.012601} {\bibfield  {journal} {\bibinfo  {journal} {Phys. Rev. A}\ }\textbf {\bibinfo {volume} {110}},\ \bibinfo {pages} {012601} (\bibinfo {year} {2024})}\BibitemShut {NoStop}%
\bibitem [{\citenamefont {Xu}\ \emph {et~al.}(2019)\citenamefont {Xu}, \citenamefont {Xie}, \citenamefont {Shi}, \citenamefont {Wang}, \citenamefont {Xu}, \citenamefont {Wang}, \citenamefont {Wang}, \citenamefont {Plenio},\ and\ \citenamefont {Du}}]{XuKebiao_ScienceAdvances_(2019)}%
  \BibitemOpen
  \bibfield  {author} {\bibinfo {author} {\bibfnamefont {K.}~\bibnamefont {Xu}}, \bibinfo {author} {\bibfnamefont {T.}~\bibnamefont {Xie}}, \bibinfo {author} {\bibfnamefont {F.}~\bibnamefont {Shi}}, \bibinfo {author} {\bibfnamefont {Z.-Y.}\ \bibnamefont {Wang}}, \bibinfo {author} {\bibfnamefont {X.}~\bibnamefont {Xu}}, \bibinfo {author} {\bibfnamefont {P.}~\bibnamefont {Wang}}, \bibinfo {author} {\bibfnamefont {Y.}~\bibnamefont {Wang}}, \bibinfo {author} {\bibfnamefont {M.~B.}\ \bibnamefont {Plenio}},\ and\ \bibinfo {author} {\bibfnamefont {J.}~\bibnamefont {Du}},\ }\bibfield  {title} {\bibinfo {title} {Breaking the quantum adiabatic speed limit by jumping along geodesics},\ }\href {https://doi.org/10.1126/sciadv.aax3800} {\bibfield  {journal} {\bibinfo  {journal} {Sci. Adv.}\ }\textbf {\bibinfo {volume} {5}},\ \bibinfo {pages} {eaax3800} (\bibinfo {year} {2019})}\BibitemShut {NoStop}%
\bibitem [{\citenamefont {Harutyunyan}\ \emph {et~al.}(2023)\citenamefont {Harutyunyan}, \citenamefont {Holweck}, \citenamefont {Sugny},\ and\ \citenamefont {Gu\'erin}}]{Harutyunyan_PRL_(2023)}%
  \BibitemOpen
  \bibfield  {author} {\bibinfo {author} {\bibfnamefont {M.}~\bibnamefont {Harutyunyan}}, \bibinfo {author} {\bibfnamefont {F.}~\bibnamefont {Holweck}}, \bibinfo {author} {\bibfnamefont {D.}~\bibnamefont {Sugny}},\ and\ \bibinfo {author} {\bibfnamefont {S.}~\bibnamefont {Gu\'erin}},\ }\bibfield  {title} {\bibinfo {title} {Digital optimal robust control},\ }\href {https://doi.org/10.1103/PhysRevLett.131.200801} {\bibfield  {journal} {\bibinfo  {journal} {Phys. Rev. Lett.}\ }\textbf {\bibinfo {volume} {131}},\ \bibinfo {pages} {200801} (\bibinfo {year} {2023})}\BibitemShut {NoStop}%
\bibitem [{\citenamefont {Stanchev}\ and\ \citenamefont {Vitanov}(2024)}]{Stanchev_PRA_2024}%
  \BibitemOpen
  \bibfield  {author} {\bibinfo {author} {\bibfnamefont {S.~G.}\ \bibnamefont {Stanchev}}\ and\ \bibinfo {author} {\bibfnamefont {N.~V.}\ \bibnamefont {Vitanov}},\ }\bibfield  {title} {\bibinfo {title} {Characterization of high-fidelity {Raman} qubit gates},\ }\href {https://doi.org/10.1103/PhysRevA.109.012605} {\bibfield  {journal} {\bibinfo  {journal} {Phys. Rev. A}\ }\textbf {\bibinfo {volume} {109}},\ \bibinfo {pages} {012605} (\bibinfo {year} {2024})}\BibitemShut {NoStop}%
\bibitem [{\citenamefont {Gevorgyan}\ and\ \citenamefont {Vitanov}(2021)}]{Gevorgyan_PRA_(2021)}%
  \BibitemOpen
  \bibfield  {author} {\bibinfo {author} {\bibfnamefont {H.~L.}\ \bibnamefont {Gevorgyan}}\ and\ \bibinfo {author} {\bibfnamefont {N.~V.}\ \bibnamefont {Vitanov}},\ }\bibfield  {title} {\bibinfo {title} {Ultrahigh-fidelity composite rotational quantum gates},\ }\href {https://doi.org/10.1103/PhysRevA.104.012609} {\bibfield  {journal} {\bibinfo  {journal} {Phys. Rev. A}\ }\textbf {\bibinfo {volume} {104}},\ \bibinfo {pages} {012609} (\bibinfo {year} {2021})}\BibitemShut {NoStop}%
\bibitem [{\citenamefont {Dridi}\ \emph {et~al.}(2020)\citenamefont {Dridi}, \citenamefont {Mejatty}, \citenamefont {Glaser},\ and\ \citenamefont {Sugny}}]{Dridi_PRA_(2020)}%
  \BibitemOpen
  \bibfield  {author} {\bibinfo {author} {\bibfnamefont {G.}~\bibnamefont {Dridi}}, \bibinfo {author} {\bibfnamefont {M.}~\bibnamefont {Mejatty}}, \bibinfo {author} {\bibfnamefont {S.~J.}\ \bibnamefont {Glaser}},\ and\ \bibinfo {author} {\bibfnamefont {D.}~\bibnamefont {Sugny}},\ }\bibfield  {title} {\bibinfo {title} {Robust control of a not gate by composite pulses},\ }\href {https://doi.org/10.1103/PhysRevA.101.012321} {\bibfield  {journal} {\bibinfo  {journal} {Phys. Rev. A}\ }\textbf {\bibinfo {volume} {101}},\ \bibinfo {pages} {012321} (\bibinfo {year} {2020})}\BibitemShut {NoStop}%
\bibitem [{\citenamefont {Zhang}\ \emph {et~al.}(2022)\citenamefont {Zhang}, \citenamefont {Liu}, \citenamefont {Shi}, \citenamefont {Song}, \citenamefont {Xia},\ and\ \citenamefont {Zheng}}]{ZhangCheng_PRA_(2024)}%
  \BibitemOpen
  \bibfield  {author} {\bibinfo {author} {\bibfnamefont {C.}~\bibnamefont {Zhang}}, \bibinfo {author} {\bibfnamefont {Y.}~\bibnamefont {Liu}}, \bibinfo {author} {\bibfnamefont {Z.-C.}\ \bibnamefont {Shi}}, \bibinfo {author} {\bibfnamefont {J.}~\bibnamefont {Song}}, \bibinfo {author} {\bibfnamefont {Y.}~\bibnamefont {Xia}},\ and\ \bibinfo {author} {\bibfnamefont {S.-B.}\ \bibnamefont {Zheng}},\ }\bibfield  {title} {\bibinfo {title} {Robust population inversion in three-level systems by composite pulses},\ }\href {https://doi.org/10.1103/PhysRevA.105.042414} {\bibfield  {journal} {\bibinfo  {journal} {Phys. Rev. A}\ }\textbf {\bibinfo {volume} {105}},\ \bibinfo {pages} {042414} (\bibinfo {year} {2022})}\BibitemShut {NoStop}%
\bibitem [{\citenamefont {Shi}\ \emph {et~al.}(2024)\citenamefont {Shi}, \citenamefont {Wang}, \citenamefont {Zhang}, \citenamefont {Song},\ and\ \citenamefont {Xia}}]{XiaYan_PRA_(2024)}%
  \BibitemOpen
  \bibfield  {author} {\bibinfo {author} {\bibfnamefont {Z.-C.}\ \bibnamefont {Shi}}, \bibinfo {author} {\bibfnamefont {J.-H.}\ \bibnamefont {Wang}}, \bibinfo {author} {\bibfnamefont {C.}~\bibnamefont {Zhang}}, \bibinfo {author} {\bibfnamefont {J.}~\bibnamefont {Song}},\ and\ \bibinfo {author} {\bibfnamefont {Y.}~\bibnamefont {Xia}},\ }\bibfield  {title} {\bibinfo {title} {Universal composite pulses for robust quantum state engineering in four-level systems},\ }\href {https://doi.org/10.1103/PhysRevA.109.022441} {\bibfield  {journal} {\bibinfo  {journal} {Phys. Rev. A}\ }\textbf {\bibinfo {volume} {109}},\ \bibinfo {pages} {022441} (\bibinfo {year} {2024})}\BibitemShut {NoStop}%
\bibitem [{\citenamefont {Mahana}\ \emph {et~al.}(2024)\citenamefont {Mahana}, \citenamefont {Davuluri},\ and\ \citenamefont {Dey}}]{Mahana_PRA_2024}%
  \BibitemOpen
  \bibfield  {author} {\bibinfo {author} {\bibfnamefont {M.~M.}\ \bibnamefont {Mahana}}, \bibinfo {author} {\bibfnamefont {S.}~\bibnamefont {Davuluri}},\ and\ \bibinfo {author} {\bibfnamefont {T.~N.}\ \bibnamefont {Dey}},\ }\bibfield  {title} {\bibinfo {title} {Coherent population transfer with polariton states in circuit \text{QED}},\ }\href {https://doi.org/10.1103/PhysRevA.110.023716} {\bibfield  {journal} {\bibinfo  {journal} {Phys. Rev. A}\ }\textbf {\bibinfo {volume} {110}},\ \bibinfo {pages} {023716} (\bibinfo {year} {2024})}\BibitemShut {NoStop}%
\bibitem [{\citenamefont {Sierant}\ \emph {et~al.}(2023)\citenamefont {Sierant}, \citenamefont {Kopciuch},\ and\ \citenamefont {Pustelny}}]{Sierant_PRA_2023}%
  \BibitemOpen
  \bibfield  {author} {\bibinfo {author} {\bibfnamefont {A.}~\bibnamefont {Sierant}}, \bibinfo {author} {\bibfnamefont {M.}~\bibnamefont {Kopciuch}},\ and\ \bibinfo {author} {\bibfnamefont {S.}~\bibnamefont {Pustelny}},\ }\bibfield  {title} {\bibinfo {title} {Tailoring population transfer between two hyperfine ground states of $^{87}\mathrm{Rb}$},\ }\href {https://doi.org/10.1103/PhysRevA.107.052810} {\bibfield  {journal} {\bibinfo  {journal} {Phys. Rev. A}\ }\textbf {\bibinfo {volume} {107}},\ \bibinfo {pages} {052810} (\bibinfo {year} {2023})}\BibitemShut {NoStop}%
\bibitem [{\citenamefont {Shapiro}\ \emph {et~al.}(2007)\citenamefont {Shapiro}, \citenamefont {Milner}, \citenamefont {Menzel-Jones},\ and\ \citenamefont {Shapiro}}]{PAP1(2007)}%
  \BibitemOpen
  \bibfield  {author} {\bibinfo {author} {\bibfnamefont {E.~A.}\ \bibnamefont {Shapiro}}, \bibinfo {author} {\bibfnamefont {V.}~\bibnamefont {Milner}}, \bibinfo {author} {\bibfnamefont {C.}~\bibnamefont {Menzel-Jones}},\ and\ \bibinfo {author} {\bibfnamefont {M.}~\bibnamefont {Shapiro}},\ }\bibfield  {title} {\bibinfo {title} {Piecewise adiabatic passage with a series of femtosecond pulses},\ }\href {https://doi.org/10.1103/PhysRevLett.99.033002} {\bibfield  {journal} {\bibinfo  {journal} {Phys. Rev. Lett.}\ }\textbf {\bibinfo {volume} {99}},\ \bibinfo {pages} {033002} (\bibinfo {year} {2007})}\BibitemShut {NoStop}%
\bibitem [{\citenamefont {Zhdanovich}\ \emph {et~al.}(2008)\citenamefont {Zhdanovich}, \citenamefont {Shapiro}, \citenamefont {Shapiro}, \citenamefont {Hepburn},\ and\ \citenamefont {Milner}}]{PAP2(2008)}%
  \BibitemOpen
  \bibfield  {author} {\bibinfo {author} {\bibfnamefont {S.}~\bibnamefont {Zhdanovich}}, \bibinfo {author} {\bibfnamefont {E.~A.}\ \bibnamefont {Shapiro}}, \bibinfo {author} {\bibfnamefont {M.}~\bibnamefont {Shapiro}}, \bibinfo {author} {\bibfnamefont {J.~W.}\ \bibnamefont {Hepburn}},\ and\ \bibinfo {author} {\bibfnamefont {V.}~\bibnamefont {Milner}},\ }\bibfield  {title} {\bibinfo {title} {Population transfer between two quantum states by piecewise chirping of femtosecond pulses: Theory and experiment},\ }\href {https://doi.org/10.1103/PhysRevLett.100.103004} {\bibfield  {journal} {\bibinfo  {journal} {Phys. Rev. Lett.}\ }\textbf {\bibinfo {volume} {100}},\ \bibinfo {pages} {103004} (\bibinfo {year} {2008})}\BibitemShut {NoStop}%
\bibitem [{\citenamefont {Wang}\ and\ \citenamefont {Zheng}(2011)}]{WangDongsheng_PRA_2011}%
  \BibitemOpen
  \bibfield  {author} {\bibinfo {author} {\bibfnamefont {D.}~\bibnamefont {Wang}}\ and\ \bibinfo {author} {\bibfnamefont {Y.}~\bibnamefont {Zheng}},\ }\bibfield  {title} {\bibinfo {title} {Quantum interference in a four-level system of a $^{87}\mathrm{Rb}$ atom: Effects of spontaneously generated coherence},\ }\href {https://doi.org/10.1103/PhysRevA.83.013810} {\bibfield  {journal} {\bibinfo  {journal} {Phys. Rev. A}\ }\textbf {\bibinfo {volume} {83}},\ \bibinfo {pages} {013810} (\bibinfo {year} {2011})}\BibitemShut {NoStop}%
\bibitem [{\citenamefont {Namazi}\ \emph {et~al.}(2017)\citenamefont {Namazi}, \citenamefont {Kupchak}, \citenamefont {Jordaan}, \citenamefont {Shahrokhshahi},\ and\ \citenamefont {Figueroa}}]{Namazi_PRAppl_2017}%
  \BibitemOpen
  \bibfield  {author} {\bibinfo {author} {\bibfnamefont {M.}~\bibnamefont {Namazi}}, \bibinfo {author} {\bibfnamefont {C.}~\bibnamefont {Kupchak}}, \bibinfo {author} {\bibfnamefont {B.}~\bibnamefont {Jordaan}}, \bibinfo {author} {\bibfnamefont {R.}~\bibnamefont {Shahrokhshahi}},\ and\ \bibinfo {author} {\bibfnamefont {E.}~\bibnamefont {Figueroa}},\ }\bibfield  {title} {\bibinfo {title} {Ultralow-noise room-temperature quantum memory for polarization qubits},\ }\href {https://doi.org/10.1103/PhysRevApplied.8.034023} {\bibfield  {journal} {\bibinfo  {journal} {Phys. Rev. Appl.}\ }\textbf {\bibinfo {volume} {8}},\ \bibinfo {pages} {034023} (\bibinfo {year} {2017})}\BibitemShut {NoStop}%
\bibitem [{\citenamefont {Kim}\ and\ \citenamefont {Marino}(2018)}]{Saesun_OE_2018}%
  \BibitemOpen
  \bibfield  {author} {\bibinfo {author} {\bibfnamefont {S.}~\bibnamefont {Kim}}\ and\ \bibinfo {author} {\bibfnamefont {A.~M.}\ \bibnamefont {Marino}},\ }\bibfield  {title} {\bibinfo {title} {Generation of $^{87}\mathrm{Rb}$ resonant bright two-mode squeezed light with four-wave mixing},\ }\href {https://doi.org/10.1364/OE.26.033366} {\bibfield  {journal} {\bibinfo  {journal} {Opt. Express}\ }\textbf {\bibinfo {volume} {26}},\ \bibinfo {pages} {33366} (\bibinfo {year} {2018})}\BibitemShut {NoStop}%
\bibitem [{\citenamefont {Sagona-Stophel}\ \emph {et~al.}(2020)\citenamefont {Sagona-Stophel}, \citenamefont {Shahrokhshahi}, \citenamefont {Jordaan}, \citenamefont {Namazi},\ and\ \citenamefont {Figueroa}}]{Sagona_PRL_2020}%
  \BibitemOpen
  \bibfield  {author} {\bibinfo {author} {\bibfnamefont {S.}~\bibnamefont {Sagona-Stophel}}, \bibinfo {author} {\bibfnamefont {R.}~\bibnamefont {Shahrokhshahi}}, \bibinfo {author} {\bibfnamefont {B.}~\bibnamefont {Jordaan}}, \bibinfo {author} {\bibfnamefont {M.}~\bibnamefont {Namazi}},\ and\ \bibinfo {author} {\bibfnamefont {E.}~\bibnamefont {Figueroa}},\ }\bibfield  {title} {\bibinfo {title} {Conditional $\ensuremath{\pi}$-phase shift of single-photon-level pulses at room temperature},\ }\href {https://doi.org/10.1103/PhysRevLett.125.243601} {\bibfield  {journal} {\bibinfo  {journal} {Phys. Rev. Lett.}\ }\textbf {\bibinfo {volume} {125}},\ \bibinfo {pages} {243601} (\bibinfo {year} {2020})}\BibitemShut {NoStop}%
\bibitem [{\citenamefont {Baksic}\ \emph {et~al.}(2016)\citenamefont {Baksic}, \citenamefont {Ribeiro},\ and\ \citenamefont {Clerk}}]{Baksic_PRL_2016}%
  \BibitemOpen
  \bibfield  {author} {\bibinfo {author} {\bibfnamefont {A.}~\bibnamefont {Baksic}}, \bibinfo {author} {\bibfnamefont {H.}~\bibnamefont {Ribeiro}},\ and\ \bibinfo {author} {\bibfnamefont {A.~A.}\ \bibnamefont {Clerk}},\ }\bibfield  {title} {\bibinfo {title} {Speeding up adiabatic quantum state transfer by using dressed states},\ }\href {https://doi.org/10.1103/PhysRevLett.116.230503} {\bibfield  {journal} {\bibinfo  {journal} {Phys. Rev. Lett.}\ }\textbf {\bibinfo {volume} {116}},\ \bibinfo {pages} {230503} (\bibinfo {year} {2016})}\BibitemShut {NoStop}%
\bibitem [{\citenamefont {Chathanathil}\ \emph {et~al.}(2023{\natexlab{a}})\citenamefont {Chathanathil}, \citenamefont {Ramaswamy}, \citenamefont {Malinovsky}, \citenamefont {Budker},\ and\ \citenamefont {Malinovskaya}}]{Chathanathil_PRA_2023}%
  \BibitemOpen
  \bibfield  {author} {\bibinfo {author} {\bibfnamefont {J.}~\bibnamefont {Chathanathil}}, \bibinfo {author} {\bibfnamefont {A.}~\bibnamefont {Ramaswamy}}, \bibinfo {author} {\bibfnamefont {V.~S.}\ \bibnamefont {Malinovsky}}, \bibinfo {author} {\bibfnamefont {D.}~\bibnamefont {Budker}},\ and\ \bibinfo {author} {\bibfnamefont {S.~A.}\ \bibnamefont {Malinovskaya}},\ }\bibfield  {title} {\bibinfo {title} {Chirped fractional stimulated \text{Raman} adiabatic passage},\ }\href {https://doi.org/10.1103/PhysRevA.108.043710} {\bibfield  {journal} {\bibinfo  {journal} {Phys. Rev. A}\ }\textbf {\bibinfo {volume} {108}},\ \bibinfo {pages} {043710} (\bibinfo {year} {2023}{\natexlab{a}})}\BibitemShut {NoStop}%
\bibitem [{\citenamefont {Chathanathil}\ \emph {et~al.}(2023{\natexlab{b}})\citenamefont {Chathanathil}, \citenamefont {Budker},\ and\ \citenamefont {Malinovskaya}}]{Chathanathil_2023_QST}%
  \BibitemOpen
  \bibfield  {author} {\bibinfo {author} {\bibfnamefont {J.}~\bibnamefont {Chathanathil}}, \bibinfo {author} {\bibfnamefont {D.}~\bibnamefont {Budker}},\ and\ \bibinfo {author} {\bibfnamefont {S.~A.}\ \bibnamefont {Malinovskaya}},\ }\bibfield  {title} {\bibinfo {title} {Quantum control via chirped coherent anti-{Stokes Raman} spectroscopy},\ }\href {https://doi.org/10.1088/2058-9565/ace3ed} {\bibfield  {journal} {\bibinfo  {journal} {Quantum Sci. Technol.}\ }\textbf {\bibinfo {volume} {8}},\ \bibinfo {pages} {045005} (\bibinfo {year} {2023}{\natexlab{b}})}\BibitemShut {NoStop}%
\bibitem [{\citenamefont {Monmayrant}\ \emph {et~al.}(2010)\citenamefont {Monmayrant}, \citenamefont {Weber},\ and\ \citenamefont {Chatel}}]{Monmayrant_JOPB_2010}%
  \BibitemOpen
  \bibfield  {author} {\bibinfo {author} {\bibfnamefont {A.}~\bibnamefont {Monmayrant}}, \bibinfo {author} {\bibfnamefont {S.}~\bibnamefont {Weber}},\ and\ \bibinfo {author} {\bibfnamefont {B.}~\bibnamefont {Chatel}},\ }\bibfield  {title} {\bibinfo {title} {A newcomer's guide to ultrashort pulse shaping and characterization},\ }\href {https://doi.org/10.1088/0953-4075/43/10/103001} {\bibfield  {journal} {\bibinfo  {journal} {J. Phys. B: At. Mol. Opt. Phys.}\ }\textbf {\bibinfo {volume} {43}},\ \bibinfo {pages} {103001} (\bibinfo {year} {2010})}\BibitemShut {NoStop}%
\bibitem [{\citenamefont {Shu}\ \emph {et~al.}(2016)\citenamefont {Shu}, \citenamefont {Ho}, \citenamefont {Xing},\ and\ \citenamefont {Rabitz}}]{ShuChuanCun_PRA_2016}%
  \BibitemOpen
  \bibfield  {author} {\bibinfo {author} {\bibfnamefont {C.-C.}\ \bibnamefont {Shu}}, \bibinfo {author} {\bibfnamefont {T.-S.}\ \bibnamefont {Ho}}, \bibinfo {author} {\bibfnamefont {X.}~\bibnamefont {Xing}},\ and\ \bibinfo {author} {\bibfnamefont {H.}~\bibnamefont {Rabitz}},\ }\bibfield  {title} {\bibinfo {title} {Frequency domain quantum optimal control under multiple constraints},\ }\href {https://doi.org/10.1103/PhysRevA.93.033417} {\bibfield  {journal} {\bibinfo  {journal} {Phys. Rev. A}\ }\textbf {\bibinfo {volume} {93}},\ \bibinfo {pages} {033417} (\bibinfo {year} {2016})}\BibitemShut {NoStop}%
\bibitem [{\citenamefont {Wilson}\ \emph {et~al.}(2018)\citenamefont {Wilson}, \citenamefont {H\'ebert}, \citenamefont {Perrella}, \citenamefont {Light}, \citenamefont {Genest}, \citenamefont {Pustelny},\ and\ \citenamefont {Luiten}}]{Wilson_PRAppl_2018}%
  \BibitemOpen
  \bibfield  {author} {\bibinfo {author} {\bibfnamefont {N.}~\bibnamefont {Wilson}}, \bibinfo {author} {\bibfnamefont {N.~B.}\ \bibnamefont {H\'ebert}}, \bibinfo {author} {\bibfnamefont {C.}~\bibnamefont {Perrella}}, \bibinfo {author} {\bibfnamefont {P.}~\bibnamefont {Light}}, \bibinfo {author} {\bibfnamefont {J.}~\bibnamefont {Genest}}, \bibinfo {author} {\bibfnamefont {S.}~\bibnamefont {Pustelny}},\ and\ \bibinfo {author} {\bibfnamefont {A.}~\bibnamefont {Luiten}},\ }\bibfield  {title} {\bibinfo {title} {Simultaneous observation of nonlinear magneto-optical rotation in the temporal and spectral domains with an electro-optic frequency comb},\ }\href {https://doi.org/10.1103/PhysRevApplied.10.034012} {\bibfield  {journal} {\bibinfo  {journal} {Phys. Rev. Appl.}\ }\textbf {\bibinfo {volume} {10}},\ \bibinfo {pages} {034012} (\bibinfo {year} {2018})}\BibitemShut {NoStop}%
\bibitem [{\citenamefont {Sobelman}(2012)}]{Sobelman2012Atomic}%
  \BibitemOpen
  \bibfield  {author} {\bibinfo {author} {\bibfnamefont {I.~I.}\ \bibnamefont {Sobelman}},\ }\href@noop {} {\emph {\bibinfo {title} {Atomic Spectra and Radiative Transitions}}},\ \bibinfo {series} {Springer Series on Atomic, Optical, and Plasma Physics}, Vol.~\bibinfo {volume} {12}\ (\bibinfo  {publisher} {Springer},\ \bibinfo {address} {New York},\ \bibinfo {year} {2012})\BibitemShut {NoStop}%
\bibitem [{\citenamefont {Steck}(2024)}]{Rb_date_2024}%
  \BibitemOpen
  \bibfield  {author} {\bibinfo {author} {\bibfnamefont {D.~A.}\ \bibnamefont {Steck}},\ }\href@noop {} {\bibinfo {title} {Rubidium {87 D Line Data}}},\ \bibinfo {howpublished} {\url{https://steck.us/alkalidata/rubidium87numbers.pdf}} (\bibinfo {year} {revision 2.3.3, 28 May 2024})\BibitemShut {NoStop}%
\bibitem [{\citenamefont {Steck}()}]{Steck2024}%
  \BibitemOpen
  \bibfield  {author} {\bibinfo {author} {\bibfnamefont {D.~A.}\ \bibnamefont {Steck}},\ }\href {http://atomoptics.uoregon.edu/~dsteck/teaching/quantum-optics/} {\emph {\bibinfo {title} {Quantum and Atom Optics}}},\ \bibinfo {note} {revision 0.16.2, 15 November 2024}\BibitemShut {NoStop}%
\bibitem [{\citenamefont {Walraven}()}]{RbBOOK(2025)}%
  \BibitemOpen
  \bibfield  {author} {\bibinfo {author} {\bibfnamefont {J.}~\bibnamefont {Walraven}},\ }\href {https://staff.fnwi.uva.nl/j.t.m.walraven/walraven/Publications_files/2025-AtomicPhysics-online.pdf} {\emph {\bibinfo {title} {Atomic Physics}}},\ \bibinfo {note} {lecture notes 2025 (Online version -unpublished)}\BibitemShut {NoStop}%
\end{thebibliography}
\end{document}